\documentclass[namedreferences]{solarphysics}
\usepackage[solaromanenum]{spr-sola-addons} 
\usepackage{graphicx}                    
\usepackage{color}                       
\usepackage[pdfborder={0 0 0 },urlcolor=blue,breaklinks]{hyperref}

\begin{document}

\begin{article}

\begin{opening}

\title{A New Calibrated Sunspot Group Series Since 1749: Statistics of Active Day Fractions}

%
\author{I.G.~\surname{Usoskin}$^{1,2}$ \sep
    G.A.~\surname{Kovaltsov} $^{3,1^\star}$\sep
    M.~\surname{Lockwood}$^{4}$\sep
    K.~\surname{Mursula}$^{1}$\sep
    M.~\surname{Owens}$^4$\sep
    S.K.~\surname{Solanki}$^{5,6}$
      }

%
\runningauthor{I.G. Usoskin \textit{et al.}}
\runningtitle{Active Day Fraction: New Sunspot Series}

%
\institute{ $^1${ReSoLVE, Space Physics Group, University of Oulu, Finland.}\\
 $^2${Sodankyl\"a Geophysical Observatory, University of Oulu, Finland.}\\
 $^3${Ioffe Physical-Technical Institute, St. Petersburg, Russia.}\\
 $^4${Department of Meteorology, University of Reading, Reading, UK}\\
 $^5${Max-Planck Institute for Solar System Research, G\"ottingen, Germany}\\
 $^6$ School of Space Research, Kyung Hee University,Yongin, Gyeonggi-Do,446-701, Korea\\
  $^\star${(visiting scientist)}\\
                     email: {Ilya.Usoskin@oulu.fi}
             }

\begin{abstract}
Although the sunspot-number series have existed since the mid-19th century, they are still the subject of intense debate,
 with the largest uncertainty being related to the ``calibration'' of the visual acuity of
 individual observers in the past.
Usually a daisy-chain regression method is applied to inter-calibrate the observers which may
 lead to significant bias and error accumulation.
Here we present a novel method to calibrate the visual acuity of the key observers to
 the reference data set of Royal Greenwich Observatory sunspot groups for the period 1900\,--\,1976, using
 the statistics of the active-day fraction.
For each observer we independently evaluate their observational thresholds [$S_{\rm S}$] defined such that the observer
 is assumed to miss all of the groups with an area smaller than $S_{\rm S}$ and report all the groups larger than $S_{\rm S}$.
Next, using a Monte-Carlo method we construct, from the reference data set, a correction matrix for each observer.
The correction matrices are significantly non-linear and cannot be approximated by a linear regression or proportionality.
We emphasize that corrections based on a linear proportionality between annually averaged data
 lead to serious biases and distortions of the data.
The correction matrices are applied to the original sunspot group records reported by the observers for each day,
 and finally the composite corrected series is produced for the period since 1748.
The corrected series is provided as supplementary material in electronic form and displays
 secular minima around 1800 (Dalton minimum)
 and 1900 (Gleissberg minimum), as well as the Modern grand maximum of activity in the second half of the 20th century.
The uniqueness of the grand maximum is confirmed for the last 250 years.
It is shown that the adoption of a linear relationship between the data of Wolf and Wolfer results
 in grossly inflated group numbers in the 18th and 19th centuries in some reconstructions.

\end{abstract}

%

\end{opening}

%

\section{Introduction}

Solar activity regularly changes in the course of the 11-year Schwabe cycle, on top of which it
 shows also slower secular variability (see, \textit{e.g.}, a review by \opencite{hathawayLR}).
This is often quantified using the sunspot number series, which covers, with different levels of
 data quality, the period since 1610 starting with the first telescopic observations.
It is generally accepted (\opencite{usoskin_LR_13}; \opencite{hathawayLR}) that solar activity varies
 between very low activity, called grand minima, such as the Maunder minimum during 1645\,--\,1715
 (\opencite{eddy76}; \opencite{usoskin_MM_15}) and grand maxima such as the recent period of high activity in the
 second half of the 20th century called the Modern Grand Maximum (\opencite{solanki_nat_04}).

The first attempt to produce a homogeneous sunspot-number series was made by R. Wolf in Zurich in the late 19th century,
 who produced the famous Wolf (sometimes also called Zurich) sunspot number series.
That effort was continued by his successors in Zurich and finally culminated at the Royal Observatory of Belgium
 as the International sunspot number series (\opencite{clette14}).
This series uses the counting method introduced by R. Wolf where the relative sunspot number [$R$] is calculated,
 for a given observer, as
\begin{equation}
R_{\rm W}=k_{\rm W}\times(10\times G + S),
\label{Eq:Rw}
\end{equation}
where $k_{\rm W}$ is the correction factor of an individual observer, and $G$ and $S$ are the numbers of sunspot groups
 and sunspots, respectively, as reported by this observer for a particular day.
Unfortunately, the raw data for this series, somewhat subjectively compiled by primary persons starting from R. Wolf,
 are not available in digital form, making it impossible to revisit except of simple corrections.

A slightly different approach was proposed by \cite{hoyt94}, who used only
 the number of sunspot groups and ignored
 the number of individual spots on the disc because their number is less robustly determined.
They formed the group sunspot number series $Rg$, where a daily value for each observer was defined as
\begin{equation}
Rg = 12.08\times k_g\times G.
\end{equation}
Here $G$ has the same meaning as in Equation~(\ref{Eq:Rw}), $k_g$ may be in general different from $k_W$,
 and 12.08 is a scaling coefficient to match the average values of $Rg$ and the Wolf sunspot number
 over the interval 1874\,--\,1976.
The original data set used by Wolf was greatly updated, nearly doubling the number of daily sunspot 
 records (\opencite{hoyt92}; \opencite{ribes93}; \opencite{hoyt95b}; \opencite{hoyt95a}; \opencite{hoyt96}) 
 culminating in the release of a full sunspot group observation
 database (referred to as HS98 henceforth \opencite{hoyt98}).
HS98 provided a comprehensive database of daily values of $G$ for all the available observers along
 with the ascribed correction factors.
This HS98 database forms a basis for further studies.
Newly recovered data are being added to it continuously, along with corrections of
 erroneous data (\textit{e.g.}, \opencite{arlt08}; \opencite{vaquero09}; \opencite{vaquero11};
 \opencite{arlt13}; \opencite{vaquero_SP_14}; \opencite{neuhauser15}).
Thus, the database of values of $G$ and $S$ exists and is kept up-to-date.
However, a major problem lies in the individual correction factors [$k$]
 (which may be different for different series) for the observers.
Since the sunspot number series is a composite series based on observations of the Sun by a large number
 of individual observers with instruments of different quality and different techniques, it is always a
 problem to produce a homogeneous series which requires an inter-calibration of the observers (\opencite{clette14}).
The standard way to ``calibrate'' observers to each other is based on a daisy chain of linear regressions
 (or even linear proportionality) between observers using periods when they overlap.
The main disadvantage of this method is that it is possible for errors to propagate and
 become accumulated over time using ``multi-store''
 regressions (a correction factor [$k$] is obtained by regression with a segment of other data,
  which has been calibrated by a regression with yet another segment, \textit{etc.}).
For example, if one observer is erroneously assessed, this error will be transferred to all other
 observers linked to that one.
The regression is typically based on daily values (\opencite{hoyt98}), but in in some cases
 (\opencite{svalgaard15} referred to as SS15 henceforth) a proportionality between heavily smoothed (annual) values
 is used which, may lead to a serious bias as shown in Section~\ref{Sec:nonlin}.

Several potential errors in the assessment of observers'' quality have been suggested recently, leading to discontinuities
 such as: the Waldmeier discontinuity in the Wolf and International sunspot
 number (\opencite{clette14}; \opencite{lockwood_1_14}) in the 1940s, related to a change in the sunspot counting algorithm at
 the Zurich observatory; a jump between observations by Wolf and by Wolfer in the 1880s (\opencite{clette14});
  and a discontinuity between
 Schwabe and Wolf data in 1848 (\opencite{leussu13}).
Thus, a need for a revision of the sunspot series by re-calibrating individual observers has become clear.

An attempt to revise the sunspot number and to produce a homogeneous data set was made recently by
 SS15 who introduced a new sunspot-group number.
The method of calibration of the observers in the 19th and 20th century and partly in the 18th century
 is a modified daisy-chain regression method.
They used several ``backbone'' key observers (Staudacher, Schwabe, Wolfer, and Koyama) so that all other observers are
 normalized (using a linear proportionality between annual values) to these ``backbones''.
Since the times at which key ``backbone'' observers (let us denote those as $A$ and $B$) carried out their
 observations do not overlap, they are normalized via ``secondary'' observers of the ``backbones'' 
 $A_1,\, A_2$, ...,  $B_1,\, B_2$, ... using linear scaling of the annual $G$-values. 
Accordingly, bridging between the core observers is performed in the following chain: 
 $A\, \leftrightarrow\, A_i\, \leftrightarrow\, B_j\, \leftrightarrow\, B$, viz. via a ``multi-store'' regression. 
Thus, comparison of the activity levels between, e.g., Staudacher and Koyama includes several subsequent regressions,
 making the whole procedure a daisy chain prone to error accumulation.]
This method has two shortcomings.
\begin{itemize}
\item
First, it is not solid \textit{viz.} while ``backbones'' may be internally solid (but see below), the connection
 between them is soft , via stretchy ``multi-store'' regressions that accumulate errors and cannot guarantee robust
 normalization.
Note that although the ``backbone'' method is said by its authors to avoid ``daisy-chaining'' this is not the case,
 since it includes calibration of data (between the backbones) derived by comparison with data from an adjacent interval
 using inter-calibration over a period of overlap between the two.
Even though they use the observer covering intermediate years to calibrate earlier
 and later years observers, the errors at either step
 propagate between the beginning and the end of thus calibrated series.
For example, the most prominent feature of the SS15 series relates to the amplitudes of solar cycles
 in the 18th and 19th centuries compared to modern cycles: these comparisons rely on a series of
  daisy-chained multi-store regressions between the older and the modern data, irrespective of the order
  in which they are carried out.
Daisy-chaining is a major concern because errors in each inter-calibration are compounded over the duration of the composite data series.
Avoiding daisy-chaining requires a calibration method that can be applied for all data segments to the same
 reference conditions, independently of the calibration of temporally adjacent data series.

\item
Second, the use of a linear proportionality between annually averaged data points is inappropriate (see Section~\ref{Sec:nonlin})
 and may lead to serious biases.
In addition, there are, in general, a great many problems and pitfalls associated with the regressions used by daisy-chaining
 (\opencite{lockwood06}; \opencite{lockwood15}).
The errors in the data can violate assumptions involved in the technique, leading to grossly misleading fits,
 even when correlation coefficients are high.
The relationship may not be linear and a regression derived for a period of low activity would inherently involve
 gross extrapolation if applied to larger-amplitude cycles (see a discussion in Section~\ref{Sec:nonlin}).
In addition, the use of the proportionality (regressions are forced through the origin) will, in general, cause amplification of
 the solar-cycle amplitudes in data from lower visual acuity observers (\opencite{lockwood15}).
\end{itemize}
Here we propose a novel method to assess the quality of individual observers and to normalize them to the
 reference data set -- the Royal Greenwich Observatory data (Greenwich Photoheliographic Results, GPR) for the
  period 1900\,--\,1976).
The method is based on comparison between the statistics of the active day fraction in the observer's data and that
 in the reference data set using pre-calculated calibration curves.
The new method allows, for the first time, totally independent calibration of each observer to a
 reference data set, without bridging them (the only exception is related to the
 data by Staudacher, where a two-step normalization is applied: see Section~\ref{Sec:Staud}).
The fact that this technique can be applied to fragments of data that are not continuous with other data
 demonstrates that the method avoids daisy-chaining and its associated error propagation:
 in the new method, if one observer is calibrated erroneously, it does not affect the other
 observers  in any way.
For each observer the observational threshold of the sunspot group size is defined such that the observer
 is assumed to miss all of the groups smaller than the threshold and report all of the larger groups.
The new method allows us to assess the quality of each observer and form a new
 homogeneous data series of the number of sunspot groups.

\section{Calibration Method}

The calibration method is based on a comparison of the statistics of the active day fraction (ADF) of the data from the observer
 in question with that of the reference data set.
The ADF (or the related fraction of the spotless days) is a very sensitive indicator of the level of solar activity around
 solar minima, more robust than the number of sunspots or groups (\opencite{harvey99}; \opencite{kovaltsov04};
  \opencite{vaquero12}; \opencite{vaquero15}).
The method includes several stages: assessing the observational quality of individual observers,
 quantified as the area of sunspot groups that they would not have noticed; recalibrating individual observers
 to the reference dataset; and compiling a composite time series.
These stages are described below.

\subsection{Reference Data Set}

We normalize all observations to the reference data set which is selected to be
 the RGO (Royal Greenwich Observatory)
 data\footnote{We use the version of the RGO data available at the Marshall Space Flight Center (MSFC)
{\sf solarscience.msfc.nasa.gov/greenwch.shtml},
 as compiled, maintained and corrected by D. Hathaway.
 This data set is slightly different
 from other versions of the RGO data stored elsewhere, \textit{e.g.}, at the National Geophysical Data Center in Boulder CO
 (\opencite{willis13b}.)}
 of sunspot groups, since it provides all the necessary information (the observed sunspot group areas) on a regular basis.
The RGO data set, often also called the Greenwich Photoheliographic Results (GPR: \opencite{baumann05}; \opencite{willis13}),
 was compiled using white-light photographs (photo-heliograms) of the Sun from a small network of observatories, giving 
 a dataset of daily observations between 17 April 1874 and the end of 1976, thereby covering nine solar cycles.
The observatories used were: The Royal Observatory, Greenwich (until 2 May 1949); the Royal Greenwich Observatory, Herstmonceux
 (3 May 1949 -– 21 December 1976); the Royal Observatory at the Cape of Good Hope, South Africa; the Dehra Dun Observatory, in the
 North-West Provinces (Uttar Pradesh) of India; the Kodaikanal Observatory, in southern India (Tamil Nadu); and the Royal Alfred Observatory in Mauritius.
Any remaining data gaps were filled using photographs from many other solar observatories, including the Mount Wilson Observatory,
 the  Harvard College Observatory, Melbourne Observatory, and the US Naval Observatory.

The sunspot areas were measured from the photographs with the aid of a large position micrometer (see \opencite{willis13}
 and references therein).
The original RGO photographic plates from 1918 onwards have survived and have been digitized by
 the Mullard Space Science Laboratory in the UK.
Automated scaling algorithms can derive sunspot areas (\opencite{cakmak14}), and it has been shown that the RGO data
 reproduce the manually scaled daily sunspot group numbers very well with a correlation of over 0.93
 (A. Tlatov and V. Ershov, private communication, 2015).
However, the RGO data may be subject to an unstable data-quality problem before 1900 (\opencite{clette14};
 \opencite{cliver15}; \opencite{willis15}).
After 1977 the RGO data have been replaced by USAF--NOAA with a different definition of sunspot-group areas (\opencite{balmaceda09}).
Accordingly, we limited the RGO reference data set to the period 1900\,--\,1976, which includes 28,644
 daily records (924 months) with full coverage.
This period includes moderate to high solar activity cycles and thus leads to a conservative upper bound
 of the observer calibration, as discussed below.
We assume that the RGO dataset corresponds to one observed by a ``perfect'' observer, who
 reports all of the sunspot groups, including the smallest one.
Although it we cannot \textit{a priori} be sure that this assumption is correct, it does not affect the calibration method.
As one can see below, we did not find observers in the 18th and 19th century the quality of whose data would be better
 than that of the RGO series.

\subsection{Assessing the Quality of Observers}
\label{S:as}

We assume that the ``quality'' of observers reporting sunspot groups is related to the size (area)
 of sunspot groups that they can see or report.
It is quantified as the threshold area [$S_{\rm S}$] (in millionths of the solar disk: msd) of the sunspot group,
 so that the observer would miss all of the groups with an area smaller than $S_S$ and observe all of the groups
 with an area greater than  $S_{\rm S}$.
We use the apparent area as seen by the observer, not corrected for foreshortening.
Here we estimate this threshold area  [$S_{\rm S}$].

\subsubsection{Calibration Curves}
\label{S:cal}

\begin{figure}[t]
\centering \resizebox{\columnwidth}{!}{\includegraphics{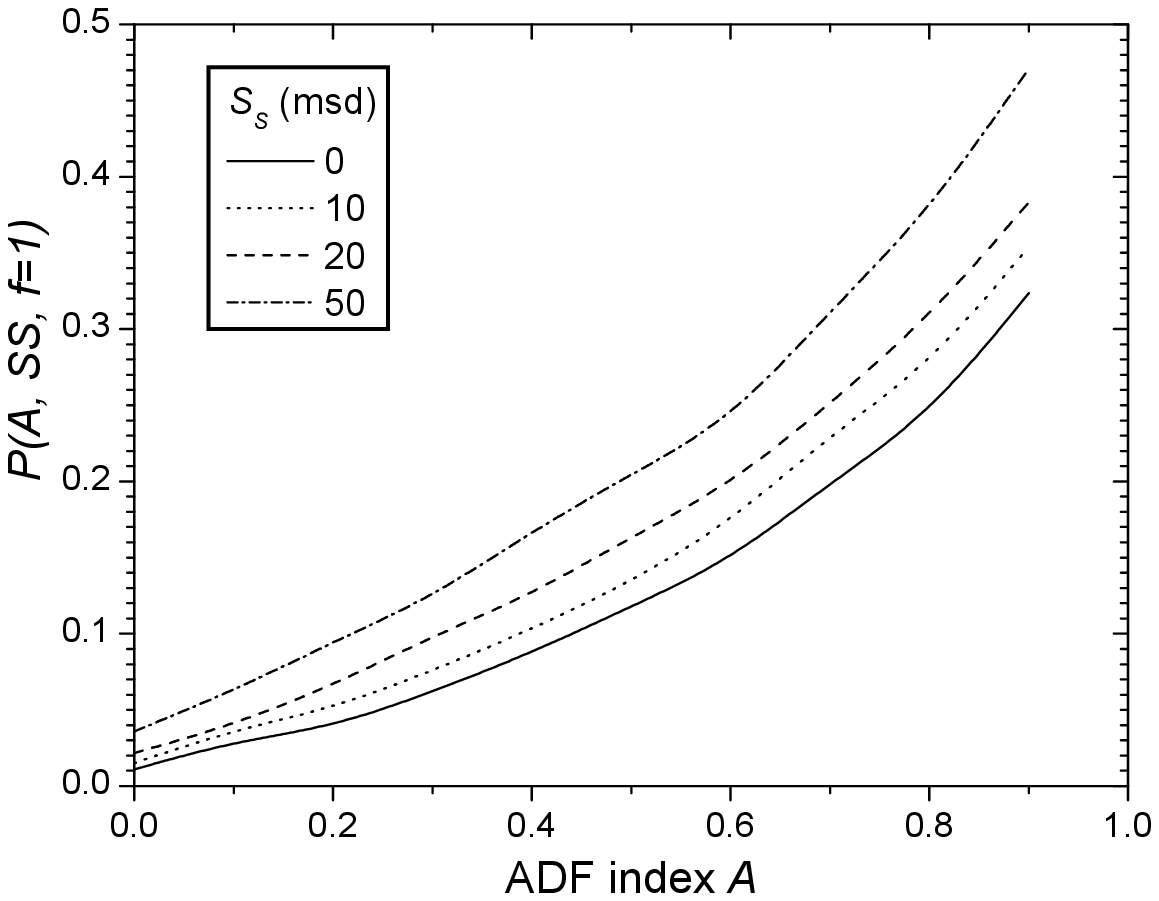}}
\caption{The cumulative distribution of $P(A*)$ (see item iii) of Section~\ref{S:as}) for the reference data set
 for different values of the threshold observed area [$S_{\rm S}$] as denoted in the legend and the data coverage
 fraction $f$=1.}
\label{Fig:cal}
\end{figure}

\begin{figure}[t]
\centering \resizebox{\columnwidth}{!}{\includegraphics{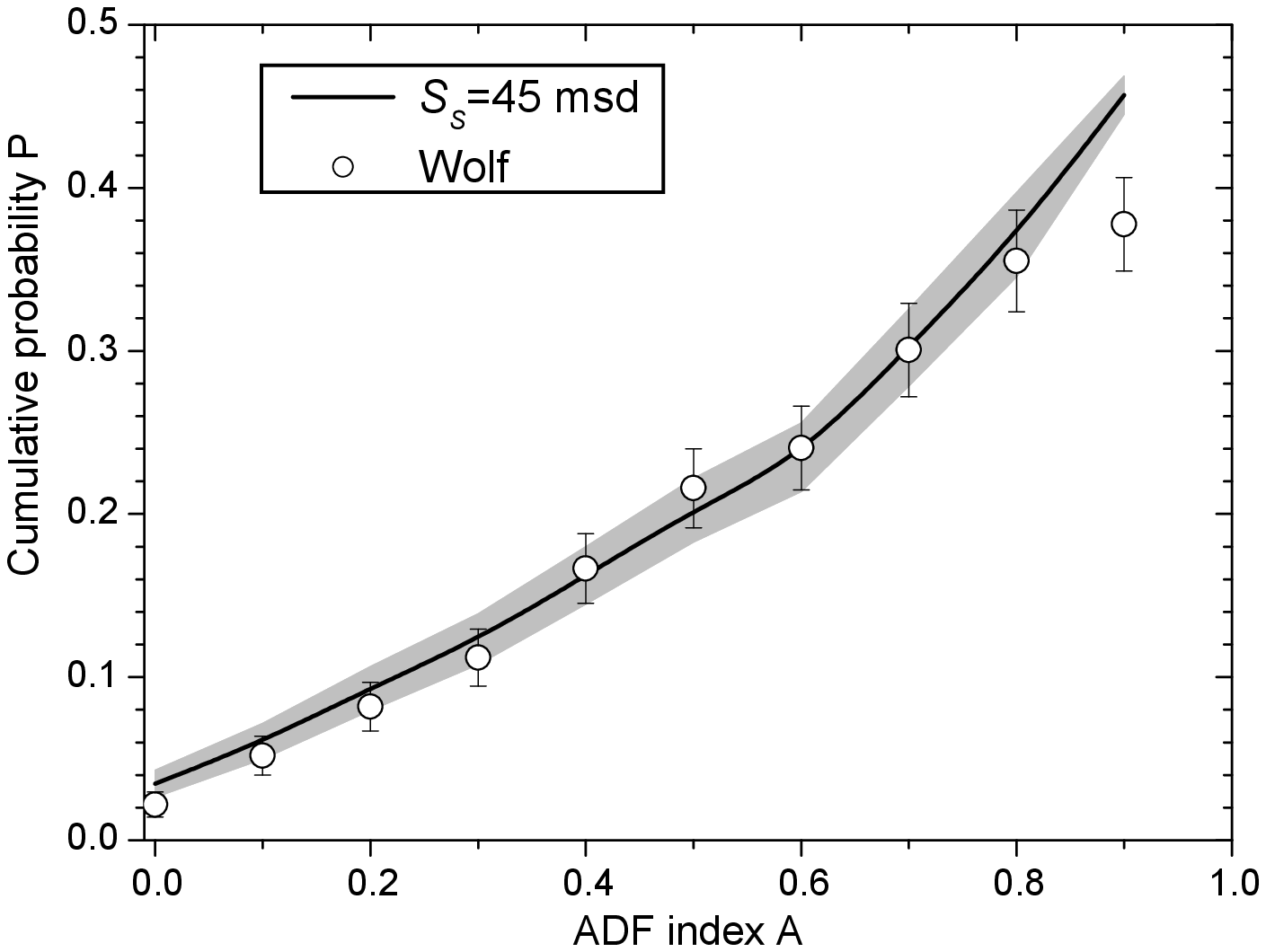}}
\caption{The cumulative probability distribution function [$P(A)$] for observational data of R. Wolf
 (circles with error bars) together with the calibration curve (\textit{cf.} Figure~\ref{Fig:cal}) for
 $S_{\rm S}=45$ msd and $f=0.66$ along with the uncertainties indicated by
 the shaded area.  }
\label{Fig:wolf}
\end{figure}

First we make calibration curves based on the reference data set to assess the quality of each observer by comparing
 to these curves.
The calibration curves were constructed by applying the following procedure:
\begin{enumerate}[i)]
\item
For each day during the reference period we counted, in the reference data set, the number of sunspot groups with
 the observed whole area exceeding a given value.

\item
For each month of the reference data set we calculated the active-day fraction (ADF) index defined
 as $A={{n_a}/n}$, where $n_a$ is the number of days with activity (at least one group is observed),
 and $n$ is the number of observational days in the month.
The ADF index [$A$] takes values from zero (no spots observed during the month) to unity (some sunspot groups observed
 for every day with observations during the month).

\item
For the whole reference period (924 months) we constructed a cumulative probability function of the ADF
 [$P(A*)= N(A\leq A*)/N$], where $A*$ is the given ADF value ranging between 0 and 0.9, $N(A\leq A*)$ is the number of
 months with the values of $A$ less than or equal to $A*$, and $N$ is the total number of months analyzed.
For example, $P(0.1)=0.029$ implies that 27 months (2.9\,\%) out of all the months in the reference data set
 have the ADF index $A\leq 0.1$
This distribution is shown in Fig.~\ref{Fig:cal} as the solid curve for $S_S=0$.
Statistical uncertainties of the $P$ values are defined as $\sigma_0(A*,S_S)=\sqrt{N(A\leq A*,S_S)/N(S_S)}$.

\item
We repeated steps 2 and 3 above, but applying an artificial observational threshold for the group area,
 \textit{viz.} $S_{\rm S}>1$, 5, 10, 15, 20, etc. (msd)
 using the reference data set.
This emulates observations of an ``imperfect'' observer who cannot see or does not report
 groups with an area smaller than the threshold value.
Distributions similar to that in item {iii)} were constructed for different values of $S_{\rm S}$,
 to form a set of calibration
 curves $P(A,S_{\rm S})$, as shown in Figure~\ref{Fig:cal}, for the range of $S_{\rm S}$ from 5 to 300 msd.

\item
Since real observers usually make observations on only a fraction $f\leq 1$ of days,
 we also emulated the effect of this and estimated the related uncertainties by a Monte-Carlo method.
For each observer we defined the fraction [$f$] for the entire period of their observation used for calibration.
We repeated steps ii)\,--\,iv) above but randomly removing the fraction (1-$f$) of daily values from the reference data set.
We did this 1000 times for each combination of $f$ and $S_{\rm S}$ and thus defined an ensemble of calibration curves $P(A,S_{\rm S},f)$.
Next we defined the mean values of $P(A,S_{\rm S},f)$ over the ensemble, which are equal within the uncertainties to $P(A,S_{\rm S},f$=1),
 and their 68\,\% uncertainties $\sigma_1(A,S_{\rm S},f)$ defined as the upper and lower 16\,\% quantiles.
The final uncertainty of the calibration curves $P(A,S_{\rm S},f)$ is defined as $\sigma(A,S_{\rm S},f)=\sqrt{\sigma_0^2+\sigma_1^2}$,
 where $\sigma_0(A,S_{\rm S},f)$ is defined similarly to step 3 above, but for the random subset of the whole
 reference dataset.
An example of the cumulative probability function $P$ is shown, along with uncertainties, in Figure~\ref{Fig:wolf}
 for $S_{\rm S}=45$ msd and $f$=0.66 as corresponding to the observations of R. Wolf from Zurich.
The value of $f$ is known from the observer's (in this case R. Wolf) records and the value of $S_{S}=45$ msd
 gives the best fit to the observed pdf for that $f$.
\end{enumerate}


Thus, a set of calibration curves [$P(A,S_{\rm S},f)$] was constructed which is used to evaluate the quality of individual
 observer's sunspot group detection as quantified in terms of ``missing'' spots with the area below the threshold $S_{\rm S}$.

\subsubsection{Assessing the Observational Quality of Individual Observers}
\label{Sec:as}
In order to assess the quality of individual observers, we compared the functions [$P$] constructed for the
 data from that observer with the calibration curves as follows.

For a given observer with the daily observational coverage fraction [$f$], we defined, similarly to step ii) in
 Section~\ref{S:cal}, the monthly ADF index [$A$] and its distribution [$P(A,f)$].
Uncertainties of the $P$-values are considered statistical, similar to those in step iii) of Section~\ref{S:cal}.
Then, for the given value of $f$ we fit the observer's distribution to the calibration curves
 [$P(A,S_{\rm S},f)$] described above
 to find the value of $S_{\rm S}$, corresponding to the observer, and its uncertainty.
The fit is done using the $\chi^2$ method for $P(A,S_{\rm S},f)$ in the range of $A$ between 0.1 and 0.8
 (eight degrees of freedom).
Small values of $A<0.1$ and higher values $A>0.8$ were not used because of low statistics and required high daily coverage
 within a month.
Moreover, definition of very small and very large values requires a large number of observational days per month,
 which may distort the corresponding statistics for observers with poor coverage.
The best-fit value of $S_{\rm S}$ is defined by minimizing the $\chi^2$-values, while uncertainties
 are defined as the range of $S_{\rm S}$ where the value of $\chi^2(S_{\rm S})$ lies below $\chi^2_0+1$ ($\chi^2_0$ being the
 minimum value) corresponding to the 68.3\,\%
 confidence interval.

\begin{figure}[t]
\centering \resizebox{\columnwidth}{!}{\includegraphics{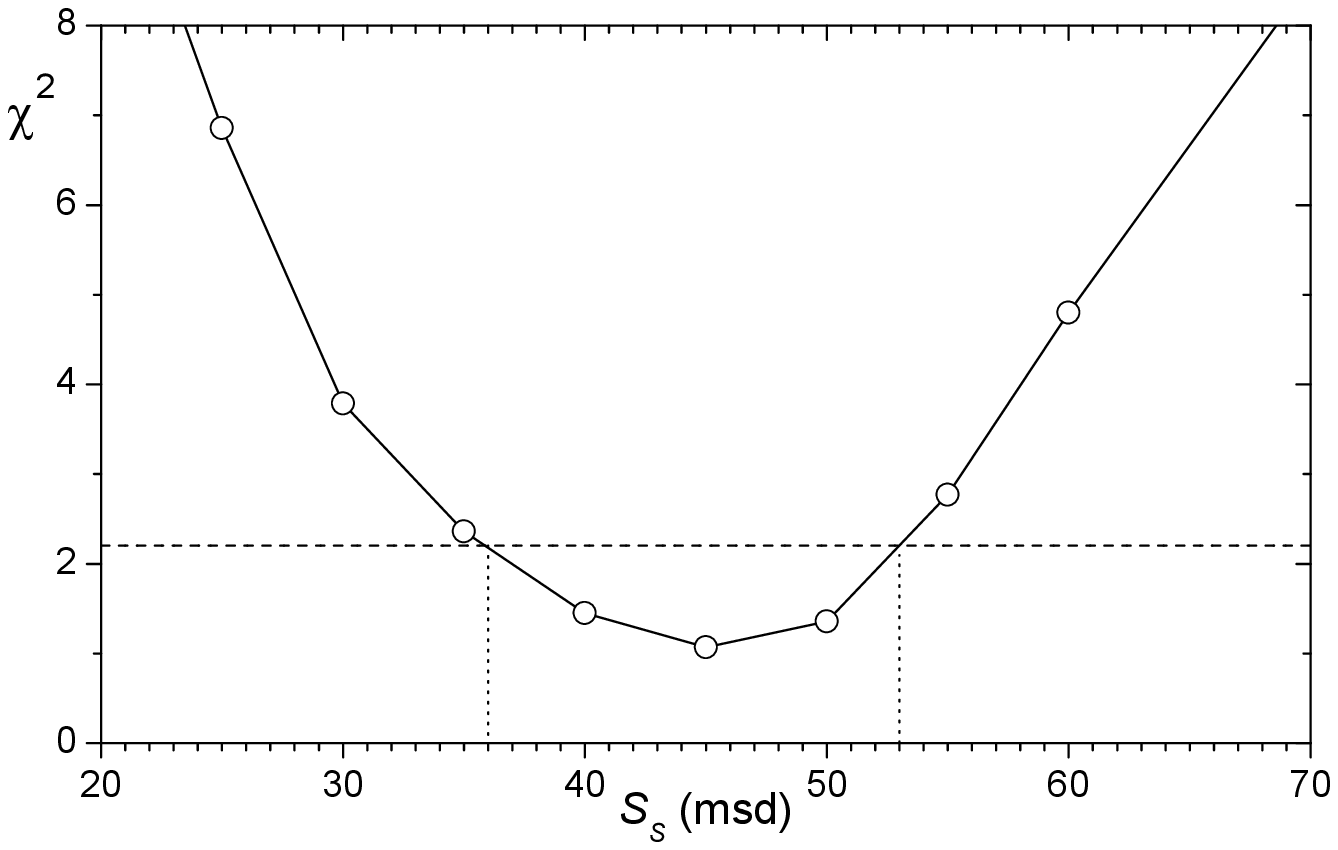}}
\caption{Dependence of the $\chi^2$ of the fitting of Wolf's data distribution to the calibration curves
 [$P(A, S_{\rm S}, f=0.66)$] on the threshold value of $S_{\rm S}$.
 The horizontal dashed and vertical dotted lines illustrate the definition of the uncertainties.}
\label{Fig:chi2}
\end{figure}

An example of the fit is shown in Figure~\ref{Fig:wolf} for observations by R. Wolf (see Table~\ref{Tab:Res}).
The $P$-distribution (dots with error bars) constructed for the data by Wolf
 is compared with the calibration curve for $S_{\rm S}=45$ msd and $f=0.66$ (the value of $f$ that applies to Wolf's data).
The dependence of the $\chi^2$-value on the value of $S_{\rm S}$ for Wolf's data is shown in Figure~\ref{Fig:chi2}.
For this particular observer the best-fit value of $S_{\rm S}$ is found to be 45 msd with the 68\,\% confidence interval being 36 to 53 msd.
This means that, on average, Wolf did not see or report sunspot groups with an (observed) area smaller than 45 msd.
We also check for a potential source of error in that, since this method compares the ADF statistics of
 the real observations to those of the reference data set,
 it may depend on the level of solar activity during observations.
If the observations were made during a period of very high activity, essentially higher than that averaged over the
 reference period, this method may yield the observer's $P$-curves to that appear lower because of
 the smaller ADF.
This may lead to a potential overestimate of the observer's ``quality'' (underestimate of their personal
 observational threshold $S_{\rm S}$) and thus to an underestimate of the solar activity based on his/her observational data.
Conversely, for low activity, the observer's curve may appear higher, leading to an underestimate
 of his/her quality (overestimate of $S_{\rm S}$) and subsequently to an overestimate of solar activity.
Since the calibration curves cover a period of moderate to high levels of activity (1900\,--\,1976), including the
 highest Cycle 19, the method overall tends to give a conservative lower limit of the observer's quality
 (upper limit of the observational thresholds [$S_{\rm S}$]).

In order to check the dependence of the $S_{\rm S}$ on the activity level, we have composed synthetic pseudo-observers
 as subsets of the RGO
 data for the periods of low activity 1902\,--\,1923 (called RGO$_{\rm low}$) and high activity 1944\,--\,1964 (RGO$_{\rm high}$).
By construction, these pseudo-observers have the true value of $S_{\rm S}=0$ since they are subsets of the reference RGO data set.
Next we calibrated these pseudo-observers using the ADF method.
We found that the formal threshold for the RGO$_{\rm low}$ observer is $S_{\rm S}$=$27\pm 10$ msd, i.e. the observer
 making observations during periods of low activity is likely to be over-calibrated (the threshold appears too high and,
 as a result, their observations, normalized to the reference data set, are overestimated).
For the RGO$_{\rm high}$ pseudo-observer we found the value of $S_{\rm S}=-16\pm 5$ msd.
The negative $S_{\rm S}$ means that we had to apply the observational threshold of 16 msd to the RGO$_{\rm high}$ data set
 in order to reproduce the ADF statistics for the reference data set with $S_{\rm S}$=0.
This would potentially lead to a slight underestimate of the corrected data for such an observer.
However, we note that the RGO$_{\rm high}$ pseudo-observer is an extreme case since the period
 around Cycle 19 was characterized by the highest activity in the entire sunspot-number series
 in all the existing sunspot series including SS15.
We did not obtained a ``negative'' threshold for any real observer considered here.
This implies that the method tends to provide an upper estimate of solar activity lying on
 the high side, particularly during periods of low-to-moderate activity.
However, the test for RGO$_{\rm low}$ shows that during the grand minima the method cannot be applied.
Future work will aim to establish calibrations for observers working during the Maunder minimum that are consistent
 with those derived here so that the data series can be extended back into the 17th century and the Maunder minimum.

We have identified 18 observers whose records can be calibrated by this method and form the core of the
 sunspot group historical series since the mid-18th century.
They are listed in Table~\ref{Tab:Res}, along with the full ranges of observational dates,
 the periods used for calibration, the spotless day fraction
 and the obtained observational threshold [$S_{\rm S}$] with its $1\sigma$ uncertainties.

For the period of the late 19th and the 20th centuries we considered only a few key observers with long stable records
 since the quality and density of data during the last hundred years were high, and thorough studies of their
 inter-calibration have been performed (\opencite{clette14}).
For the observers Quimby and Wolfer who have a significant overlap with the reference data set, we used the
 overlap period for a direct calibration.

For the period before the mid-19th century, all of the observers were considered and calibrated whenever possible
 following the method described here (i.e. whenever they had sufficient observations to apply our technique).
For each observer we used data of sunspot group counts from the HS98 database (\opencite{hoyt98}) except for
 Schwabe and Staudacher (see comments below).
Whenever possible we used data for complete solar cycles.
Unless indicated otherwise, we considered for each observer only months with three or more daily observations.

We also show in Table~\ref{Tab:Res} the spotless day fraction (SDF) which is the number of days
 with the reported absence of spots to the total number of observational days during the calibration period $T_{\rm cal}$,
 for each observer.
The reference RGO data set contains $\approx 16$\,\% of spotless days.
SDF is in the range of 15\,\% to 26\,\% for most of the observers in the 19th century except of Stark and Derfflinger
 whose observations were likely reflecting the low activity around the Dalton minimum.
Apparently different from all other was Staudacher with only 57 spotless days (5.5\,\%) reported.
He likely was more interested in drawing spots than reporting their absence.
Whatever was the reason, his statistic of spotless/active days is distorted and cannot be used to calibrate
 his quality directly to the reference period, as discussed in Sect.~\ref{Sec:Staud}.

Some specific comments are given below.

{\it S.H. Schwabe} observed the Sun during 1825\,--\,1867; however we considered for calibration only
 the period of 1832\,--\,1867, covering three full cycles, since the earlier part of his record is thought to be of
  less stable quality (\opencite{leussu13}).
Sunspot group numbers for Schwabe's observations were taken not from the HS98 database but from a new
 revised collection by
 \cite{arlt13} (available at {\sf www.aip.de/Members/rarlt/sunspots/schwabe} as version 1.3 from 12 August 2015).

For {\it J.W. Pastorff} from Drossen we used data for the observer \# 263 in the HS98 database.
His SDF is low (14\,\% -- see Table~\ref{Tab:Res}) indicating that he might skipped reporting some spotless days.
If this is true, it may lead to an overestimate of his observational quality and consequently to an underestimate of
 the sunspot group number based on his record.
On the other hand, his corrected data is consistent with those of Schwabe and Stark (Figure~\ref{Fig:all_m}).

{\it C. Horrebow} from Copenhagen (observer \#180 in the HS98 database) observed the Sun during 1761\,--\,1776.
Here we use, for calibration, the period of 1766\,--\,1776 (one solar cycle) because the data are very sparse before 1766.

{\it T. Derfflinger} from Kremsm\"unster observed the Sun during 1802\,--\,1824, but we used for calibration data covering
 1816\,--\,1824 to exclude the Dalton minimum so as to avoid the potential problems discussed above associated with a
 mismatch in the range of the data compared to that for the reference data set.

{\it J.M. Stark} from Augsburg observed the Sun during 1813\,--\,1836, and we use all data for the observer \#255
 of the HS98 database, while not considering the generic no-sunspot day records
 (observer \#254 called ``STARK, AUGSBURG, ZERO DAYS'' in the HS98 database).

{\it J.C. Schubert} from Danzig observed the Sun during 1754\,--\,1758.
We used for calibration the period of 1754\,--\,1757 which is $\pm2$ years around the formal
 cycle minimum in 1755.2.
This includes 404 daily observations with 36\,\% coverage.
To assess the quality of Schubert's observations we used the RGO statistics, as described in Section~\ref{S:cal}, but
 using RGO data only within $\pm$two years around the solar cycle minima to be consistent with Schubert's cycle coverage.

{\it J.C. Staudacher} from N\"urnberg, while providing about 1035 daily drawings for the period 1749\,--\,1795
 ($\approx$6\,\% coverage), as published by \cite{arlt08}, cannot be directly calibrated in the way proposed here
 because he did not properly report days without sunspots.
He reported no spots for only 5.5\,\% days which is less than all other observers (15\,--\,40\,\%, see Table~\ref{Tab:Res}).
Moreover, there is no zero-spot months (with the number of daily observations more than two) in his record, in contrast
 to all other observers.
This distorts the ADF statistics, making impossible direct calibration as described above.
The case of Staudacher is considered separately in Section~\ref{Sec:Staud}.

The following observers produced sufficiently long observational records but cannot be calibrated
 in the manner described above because of sparse or
 unevenly distributed observations or because they did not report spotless days: Lindener, Tevel, Arago, Heinrich, Flaugergues, Hussey.
We also did not consider observers during and around the Maunder minimum because of the very low level of activity
 (\opencite{usoskin_MM_15}) when the method cannot be applied.
The period between the end of the Maunder minimum and the mid-18th century cannot be studied because of a lack of
 sufficient observations (\opencite{vaquero09}).

\begin{table}
\caption{Results of calibration of the key observers used here. Columns are: Name of the observer;
 Period of observation $T_{\rm obs}$; Period used for calibration $T_{\rm cal}$; Number of observational days $N$ used for
 calibration; Data coverage in \,\% $f$; The fraction of spotless days (SDF) in the record;
 The threshold area $S_{\rm S}$ in uncorrected msd; values in parentheses denote the upper and lower 1$\sigma$ bound.
For details see text. }
\begin{tabular}{lcccccc}
\hline
Observer & $T_{\rm obs}$ & $T_{\rm cal}$ & $N$& $f$ &SDF & $S_{\rm S}$\\
\hline
RGO & 1874\,--\,1976 & 1900\,--\,1976 & 28644 & $\approx 100\,\%$ & 16\,\% & 0 \\
Quimby & 1889\,--\,1921 & 1900\,--\,1921$^c$ & 10830 & 92\,\% & 23\,\% &22$\left(^{28}_{16}\right)$ \\
Wolfer & 1876\,--\,1928 & 1900\,--\,1928$^c$ & 7165 & 68\,\% & 21\,\% &6$\left(^{12}_{\, 0}\right)$ \\
Winkler$^a$ & 1882\,--\,1910 & 1889\,--\,1910$^d$ & 4812 & 60\,\% & 24\,\% & 53$\left(^{66}_{45}\right)$ \\
Tacchini & 1871\,--\,1900 & 1879\,--\,1900$^d$ & 6235 & 78\,\% & 19\,\% &10$\left(^{14}_{\, 7}\right)$ \\
Leppig & 1867\,--\,1881 & 1867\,--\,1880$^d$ & 2463 & 52\,\% & 26\,\% & 45$\left(^{33}_{55}\right)$ \\
Spoerer & 1861\,--\,1893 & 1865\,--\,1893$^d$ & 5386 & 53\,\% & 15\,\% & 3$\left(^{5}_{0}\right)$ \\
Weber & 1859\,--\,1883 & 1859\,--\,1883 & 6981 & 79\,\% & 19\,\% &22$\left(^{28}_{16}\right)$\\
Wolf & 1848\,--\,1893 & 1860\,--\,1893$^d$ & 8102 & 66\,\% & 21\,\% &45$\left(^{53}_{36}\right)$ \\
Shea & 1847\,--\,1866 & 1847\,--\,1866 & 5538 & 79\,\% & 20\,\% &25$\left(^{33}_{18}\right)$\\
Schmidt & 1841\,--\,1883 & 1841\,--\,1883 & 6887 & 49\,\% & 21\,\% &10$\left(^{15}_{\,\,6}\right)$ \\
Schwabe & 1825\,--\,1867 & 1832\,--\,1867 & 8570 & 65\,\% & 18\,\% &13$\left(^{18}_{\,\,8}\right)$ \\
Pastorff & 1819\,--\,1833 & 1824\,--\,1833$^d$ & 1451 & 41\,\% & 14\,\% &5$\left(^{10}_{0}\right)$ \\
Stark$^a$ & 1813\,--\,1836 & 1813\,--\,1836$^e$ & 2406 &30\,\% & 41\,\% &60$\left(^{70}_{50}\right)$ \\
Derfflinger$^a$ & 1802\,--\,1824 & 1816\,--\,1824$^e$ & 346 & 11\,\% & 38\,\% &50$\left(^{80}_{40}\right)$ \\
Herschel$^a$ & 1794\,--\,1818 & 1795\,--\,1815 & 344 & 4\,\% & 16\,\% &23$\left(^{35}_{10}\right)$ \\
Horrebow & 1761\,--\,1776 & 1766\,--\,1776$^e$ & 1365 & 34\,\% & 27\,\% &75$\left(^{95}_{60}\right)$ \\
Schubert & 1754\,--\,1758 & 1754\,--\,1757$^e$ & 404 & 36\,\% & 37\,\% &10$\left(^{16}_{\,5}\right)$ \\
Staudacher$^b$ & 1749\,--\,1799 & 1761\,--\,1776 & 1035 & 14\,\% & 5.5\,\% & -- \\
\hline
\end{tabular}
\label{Tab:Res}
\\Notes: $^a$ -- the observational threshold [$S_{\rm S}$] is likely overestimated.\\
$^b$ -- calibration is done via Horrebow (see Section~\ref{Sec:Staud}).\\
$^c$ -- direct overlap with RGO data set.\\
$^d$ -- to use complete cycles.\\
$^e$ -- see comments in Section~\ref{Sec:as}.\\
\end{table}

In addition we also checked the record by H. Koyama from Tokyo who observed the Sun over the period 1947\,--\,1984
 with about 56\,\% daily coverage, data which formed one ``backbone'' for the method by SS15.
We used for calibration the period of 1953\,--\,1976 to cover full cycles and to be more consistent with the RGO time interval,
A total of 4778 daily observations were processed.
The calibration was performed using the reference RGO dataset for the same period of time.
The threshold [$S_{\rm S}$] value was found to be $8\pm 5$ msd, yielding a result fully consistent
 with the RGO data (see Figure~\ref{Fig:all_m}).
We stress that this record was not used in the compilation of the final series, but only to test the method.

The calibration method works after 1754 when Schubert started observing.
If Staudacher's data is included (see Section~\ref{Sec:Staud}), the calibration starts in 1749.
Before Staudacher there is a paucity of sufficiently long timeseries of observations by single observers,
 so that the method cannot work due to too poor statistics, and before 1715 the method is not applicable
 because of the Maunder minimum where the statistics of the reference data set cannot be applied.
We stopped the calibration in 1900 since the reference data set of RGO data is used after 1900.

\subsection{Corrections of Individual Observers}
\label{Sec:corr}
Once the observational threshold 8$S_{\rm S}$] and its uncertainty are defined for each observer (see above),
 observations (the number of sunspot groups) by this observer can be calibrated to the reference data set.
All corrections are done at the daily scale, because of the non-linearity of correction that may otherwise
 distort the relation, as discussed below.

\begin{figure}[t]
\centering \resizebox{\columnwidth}{!}{\includegraphics{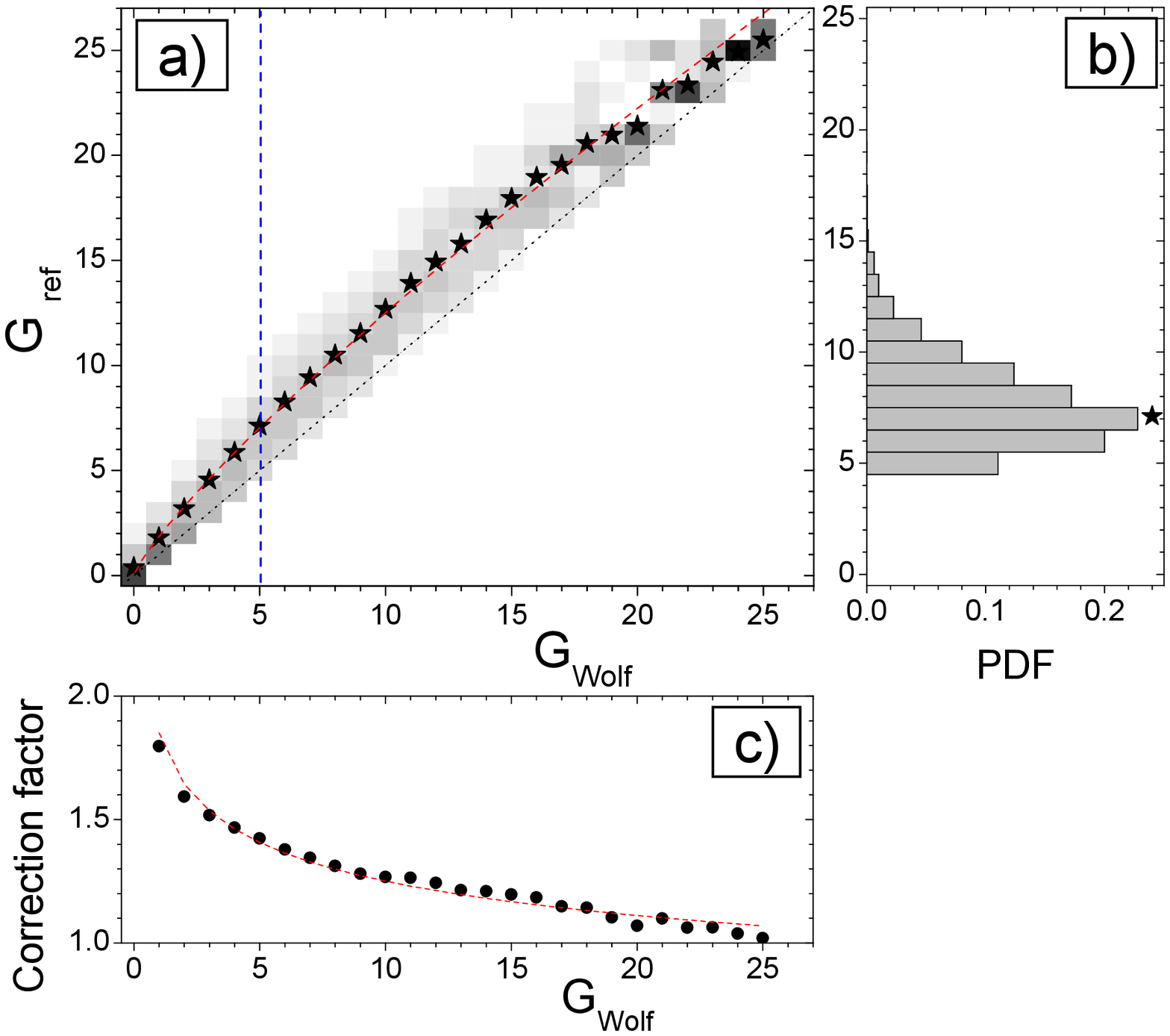}}
\caption{a: Correction matrix for the data by R. Wolf:
Distribution of the daily number of groups in the reference data set $G_{\rm ref}$ (ordinate) as
 a function of the number of groups $G_{\rm Wolf}$ reported by R. Wolf (abscissa).
 Each vertical strip is a probability density function (pdf).
 The grey scale is linear from 0 (white) to 1 (black)
 Stars represent the mean value of $G_{\rm ref}$ for the given $G_{\rm Wolf}$.
 The dotted line is the diagonal ($G_{\rm Wolf}=G_{\rm ref}$), and the dashed-red curve is the
  best-fit power law ($G_{\rm Wolfer}=1.85\cdot G_{\rm Wolf}^{0.83}$).
 Panel b: a cross-section of panel a at $G_{\rm Wolf}=5$ (vertical blue dashed line), the star denotes the mean
  value of the pdf (in this case at 7.12, making the optimum correction factor 7.12/5 = 1.424).
 The median at this example $G_{\rm Wolf}=5$ is $G_{\rm ref}=7.65$.
 Panel c: The optimum correction factor, viz. the ratio of the mean $G_{\rm ref}$ to $G_{\rm Wolf}$.
 The red-dashed line corresponds to the best-fit power law (see panel a).
 }
\label{Fig:pdf}
\end{figure}

The correction, for a given range of $S_{\rm S}$-values (see Table~\ref{Tab:Res}), is made by the Monte-Carlo
 method using the reference data set of daily RGO group numbers in the following steps:
\begin{enumerate}[i)]
\item
A test value $S_{\rm S}^*$ is randomly selected, using the normally distributed random numbers, from the
 distribution of $S_{\rm S}$ values for the observer (see Table~\ref{Tab:Res}).
A ``degraded'' subset of the reference daily data set is constructed by considering only sunspot groups with the (uncorrected)
 area $\geq S_{\rm S}^*$, \textit{i.e.} what an observer with the observational limit of $S_{\rm S}^*$ would have recorded.
For each daily value $G_{S_{\rm S}^*}$ from the degraded data set we construct a distribution of the $G_{\rm ref}$ values
 from the reference data set ($S_{\rm S}$=0), similar to that shown in Figure~\ref{Fig:pdf}b.

\item
Step 1 above is repeated 1000 times, each time randomly selecting an $S_{\rm S}^*$ value for a given observer,
 and summing up all of the distributions of $G_{\rm ref}$ for a given $G_{S_{\rm S}^*}$
The probability density function (pdf) of the $G_{\rm ref}$ values for each $G_{S_{\rm S}^*}$ value
 (\textit{i.e.} reported by the observer)
 is constructed.
Finally, the correction matrix for the particular observer is constructed as illustrated in Figure~\ref{Fig:pdf}a.

\item
For each daily recorded value [$G$] of the observer, the corresponding mean and the 68\,\% upper and lower
 quantiles of the $G_{\rm ref}$ were calculated, giving the mean corrected daily group number and its uncertainties.
\end{enumerate}

The method is illustrated in Figure~\ref{Fig:pdf} by calibration of the observational quality of R. Wolf from Zurich.
The correction of an ``imperfect'' observer (R. Wolf in this example) is based on an assessment of how many sunspot groups
 the ``perfect'' observer (the reference RGO in our case) would see for a day when the ``imperfect'' observer reported
 $G_{\rm Wolf}$ groups.
Thus, for a given $G_{\rm Wolf}$ value ($x$-axis) one obtains a pdf of the reference values [$G_{\rm ref}$] to yield
 the mean and the uncertainties of the corrected number of sunspot groups.
We note that the relation is well-approximated by a power law with the spectral index 0.83 (the dashed red curve in Figure~\ref{Fig:pdf}),
 but this functional form is shown only for illustration and not used in the construction of correction matrices.
An important feature observed is that the mean $G_{\rm ref}$ value is non-zero (0.38) for spotless days reported by
 an ``imperfect'' observer, R.Wolf in our example.
Accordingly, we cannot say whether zero spots by Wolf implies a true spotless day or whether groups
 were small and went undetected.
It is important that the correction implied by the matrix cannot be approximated by a linear (or worse still, a proportional) regression.
Panel c of Figure~\ref{Fig:pdf} depicts the daily correction factor defined as the ratio of the $G_{\rm ref}$ (the number of groups
 the real observer would see if they would be a perfect observer) to $G_{\rm Wolf}$ (the number of groups the observer R. Wolf
 actually reported).
The ratio gradually drops from $\approx 1.8$ for one group reported by Wolf to 1.19 for 15 groups reported by Wolf.
If Wolf saw 25 groups, the correction would have been only 1.02.
This implies that the larger the number of sunspot groups is (the higher the activity is), the smaller
 is the relative error of the ``imperfect'' observer.
It is clear that a simple linear regression cannot be used to correct an ``imperfect'' observer (see Section~\ref{Sec:nonlin}).
This was also emphasized by \cite{lockwood15} from a study of the effects of imposing different
 observational thresholds on the RGO data.

\begin{figure}[t]
\centering \resizebox{\columnwidth}{!}{\includegraphics{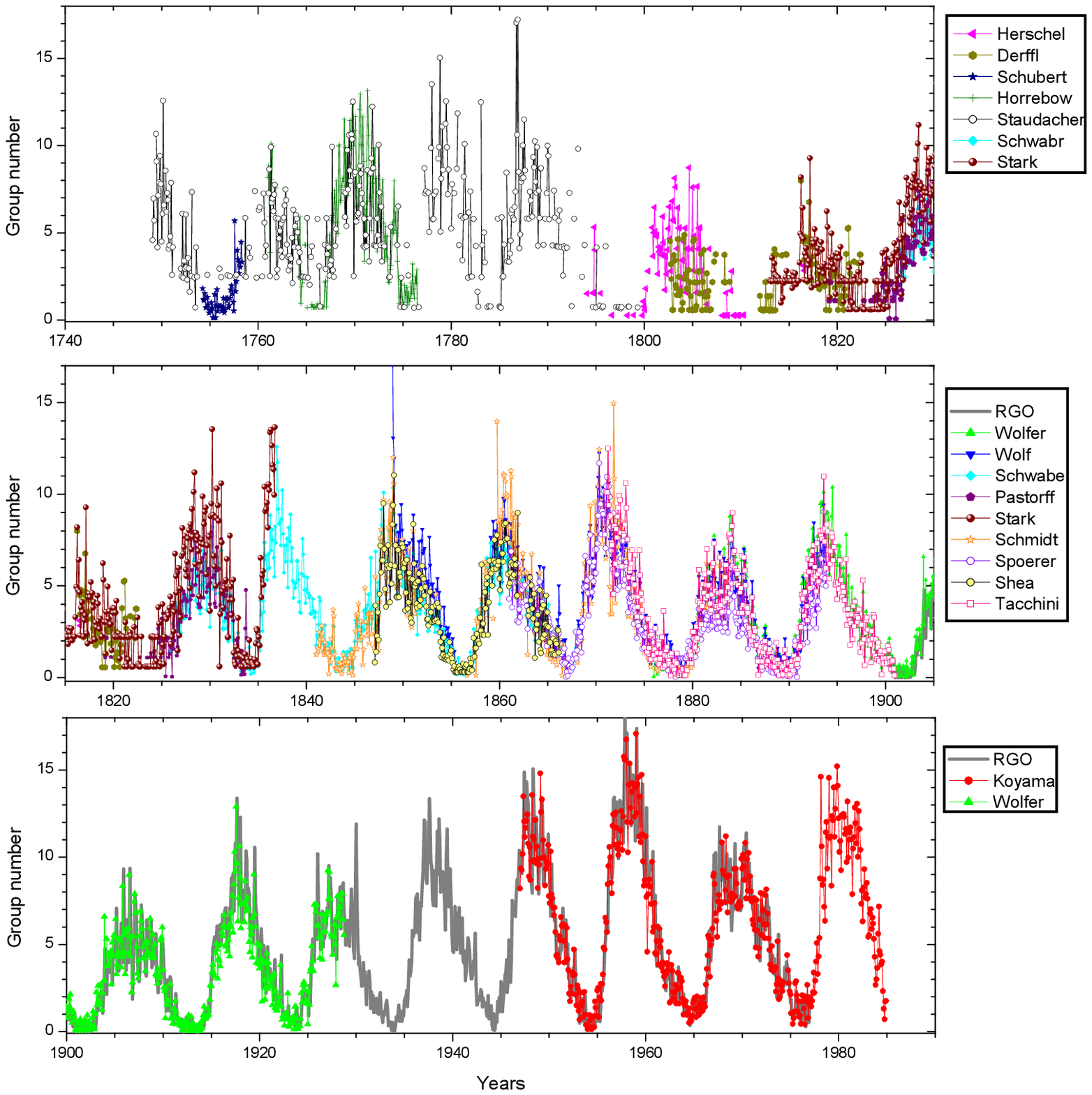}}
\caption{Monthly series of sunspot group numbers obtained by individual observers, after corrections for the observational ``imperfectness''
 (uncertainties are nor shown).}
\label{Fig:all_m}
\end{figure}
From daily values for each observer, corrected for the ``imperfectness'' as described above, we have calculated monthly
 $G_{\rm ref}$ values for individual observers, as a weighted (with weight being inversely proportional to the squared error
 of the daily corrected value) average of the available daily values.
We stress again that the correction should be applied to the daily values, not to monthly, or even worse,
 annual averages, because of the nonlinearity of the correction (see Section~\ref{Sec:nonlin}).
Such series of the monthly $G_{\rm ref}$, scaled to the reference data set of RGO, are shown in Figure~\ref{Fig:all_m}
 for some individual observers.
One can see that the different observers, after correction, agree well with one another, even though
 the corrections were done totally independently for each observer.

\subsubsection{Calibration of Staudacher via Horrebow}
\label{Sec:Staud}

J.C. Staudacher is a key observer to evaluate solar activity in the second half of the 18th century
 and a ``backbone'' observer for SS15.
It is crucially important to evaluate the quality of the data he produced.
However, since he apparently did not properly (see Table~\ref{Tab:Res}) report spotless days, being primarily interested in
 drawing sunspots, the ADF method used here cannot be directly applied to his data,
 and it would yield an unrealistically high quality of observations.
Accordingly, we have made a two-step calibration of Staudacher data to the reference data set.
This is the only exception to the method described above.

\begin{figure}[t]
\centering \resizebox{\columnwidth}{!}{\includegraphics{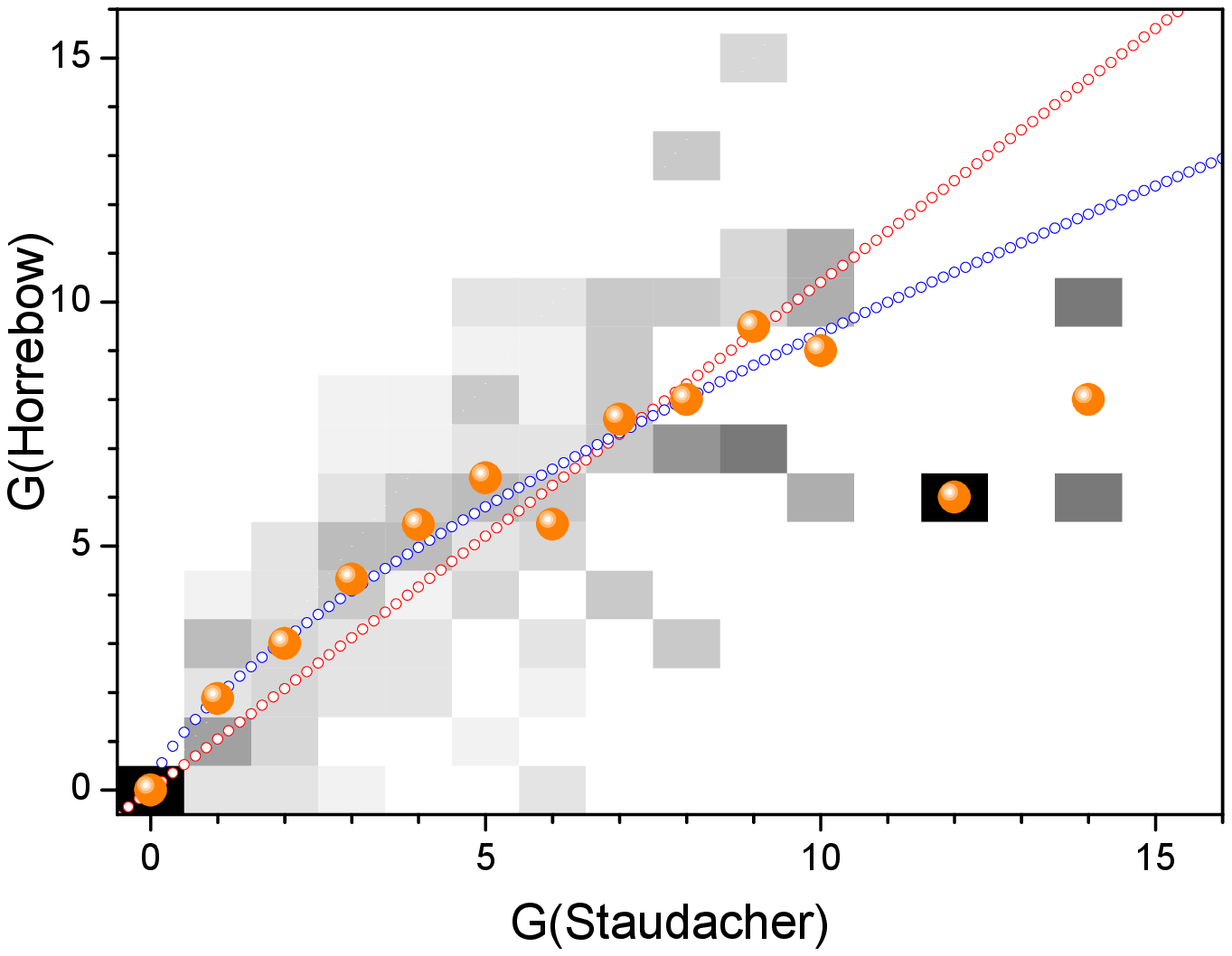}}
\caption{The pdf matrix of conversion of daily group numbers between Staudacher
 and Horrebow ($G_{\rm Staud}$ and $G_{\rm Horr}$, respectively) using direct data for overlapping (within $\pm$two days) days.
 The grey scale is linear from 0 (white) to 1 (black).
 Orange balls are the mean values.
 For illustration the best-fit linear proportionality ($G_{\rm Horr}=1.04\cdot G_{\rm Staud}$ red line) and power-law ($G_{\rm Horr}=1.91\cdot G_{\rm Staud}^{0.69}$)
  are shown.
 The two last points were not used in the analysis.
 }
\label{Fig:Staud}
\end{figure}

First we directly calibrated Staudacher's data to those recorded by C. Horrebow from Copenhagen (observer \# 180 of
 the HS98 database), whose quality is evaluated by the ADF method (Table~\ref{Tab:Res}).
We used the period of direct overlap of the two observers in 1761\,--\,1776.
Staudacher's data were digitized recently by \cite{arlt08} who processed his original drawings.
The sunspot groups were redefined (R. Arlt, personal communication 2015) using these drawings
 (\opencite{senthamizh16}).
We note that this dataset is different from the HS98 database, in particular in that it yields $\approx 30$\,\% more sunspot groups.
Horrebow's data were taken from the HS98 database as being robustly defined (R. Arlt, personal communication, 2015).
We found 110 days when both observers reported observations.
In order to improve statistics we compared also the neighboring $\pm$two days.
This added 101 cases when the observations were in successive days, and 39 days when they were separated by two days.
We note that, as discussed by \cite{willis15} in their survey of the early RGO data,
 there are some sunspot groups that last for only one day, but they are infrequent
 and likely near or below the $S_{\rm S}$ threshold for Staudacher and Horrebow, anyway.
Accordingly, the possible error introduced by using neighboring days is greatly outweighed by the reduction in uncertainty
 brought about by having a greater number of samples, more than doubling the statistics.
First we checked, for each daily observation by Staudacher, if there was a coincident record from Horrebow
 and took this pair if it existed in the positive case.
If such a counterpart was not found, we looked for an adjacent ($\pm$one day) observation by Horrebow
 and took that pair if it existed.
Otherwise, we considered the $\pm$two days interval.
This gives in total 250 pairs of ``coincident'' data days by Staudacher and Horrebow.
A cross-matrix of such daily values Staudacher \textit{vs} Horrebow was constructed as shown in Figure~\ref{Fig:Staud}.
One can see that the two observers are close to each other in the quality of observations (the matrix is
 nearly diagonal) except for the two right-most points, where Staudacher drew significantly more groups than
 reported by Horrebow.
A matrix for converting Staudacher's daily values of $G$ into Horrebow's values is constructed in a similar way
 to those above for all other observers with respect to the reference data set.

Next, we applied the correction matrix of Horrebow-to-reference data set and finally converted Staudacher's daily
 number of groups into the reference data set as follows:
\begin{equation}
\nonumber
G_{\rm Staud} \stackrel{R_1}{\longrightarrow} G^{'}_{\rm Horr} \stackrel{R_2}{\longrightarrow} G_{\rm ref}
\label{Eq:a}
\end{equation}
For each daily value $G_{\rm Staud}$ we randomly selected (step $R_1$) the value of $G^{'}_{\rm Horr}$ from the
 matrix described above (\textit{i.e.} randomly from all values corresponding to $G_{\rm Staud}$), and then, for
  this $G^{'}_{\rm Horr}$ we randomly selected a value of $G_{\rm ref}$ from the Horrebow correction matrix
  (from within the row corresponding to $G^{'}_{\rm Horr}$).
This procedure was repeated 1000 times, and finally the correction matrix of Staudacher [$G_{\rm Staud}$] daily values
 to the reference data set [$G_{\rm ref}$] was constructed.

\subsubsection{Test of the Correction: Wolf-vs-Wolfer}

\begin{figure}[t]
\centering \resizebox{10cm}{!}{\includegraphics{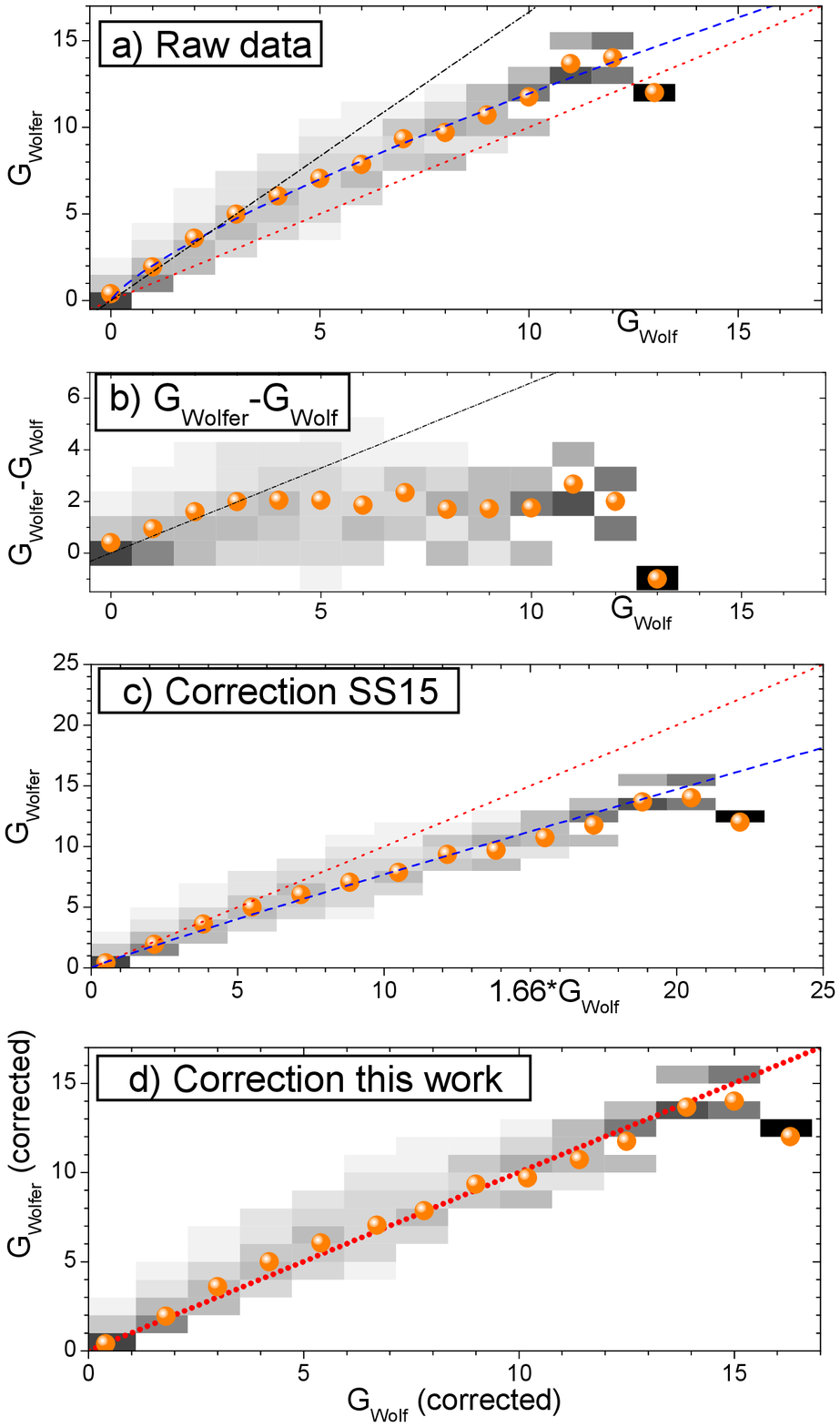}}
\caption{Comparison of daily group numbers by Wolfer and Wolf for days when both have records
 (4385 days of simultaneous records), presented as a pdf (grey scale from 0 (white) to 1 (black)).
 The orange balls represent the mean values of $G_{\rm Wolfer}$ in each bin of $G_{\rm Wolf}$ values.
 The red line depicts the diagonal ($G_{\rm Wolf}=G_{\rm Wolf}$).
 Panel a presents raw (uncorrected) daily group numbers in the HS98 database.
 The blue dashed line is the best-fit power law ($G_{\rm Wolfer}=2.03\cdot G_{\rm Wolf}^{0.77}$),
 the black dash-dotted line is the relation $G_{\rm Wolfer}=1.66\times G_{\rm Wolf}$ (used by Clette et al. (2014) and SS15).
 Panel b depicts the distribution of the difference between $G_{\rm Wolfer}$ and $G_{\rm Wolf}$ daily
  group numbers as a function of $G_{\rm Wolf}$.
 Panel c presents data by Wolf multiplied by 1.66 as proposed by SS15.
 Blue dashed curve is the best-fit power law ($G_{\rm Wolfer}=0.89\cdot G_{\rm Wolf}^{0.94}$),
 Panel d presents data of both observers corrected to the reference conditions as proposed in this work. }
\label{Fig:wolf_wolfer}
\end{figure}

A good and important example to test our and other calibration methods is the relation between data
 reported by J.R. (Rudolf) Wolf and H.A. (Alfred) Wolfer, both from the Zurich observatory.
First, a proper comparison of the two observers is crucially important as Wolf was the reference observer for the Wolf
 sunspot number series,
 while Wolfer is the reference observer for the ISN (v.2) and a ``backbone'' observer for the SS15 reconstruction.
Without an adequate comparison between them, it is difficult to compare the various sunspot series now available.
Second, a long period of their overlap (4385 days during 1876\,--\,1993) exists when both observers independently
 reported sunspots.
HS98, using a linear regression applied to the daily values of the overlapping periods, proposed that the two observers
 are quite close to each other in the quality of their observations.
In contrast, SS15 proposed, using a linear-regression analysis of the annually averaged number of sunspot
 groups, that the relation between
 them is a linear and proportional scaling so that $G_{\rm Wolfer}=1.66\times G_{\rm Wolf}$.

Figure~\ref{Fig:wolf_wolfer}a shows the scatter plot (in the form of a pdf as in Figure~\ref{Fig:pdf}a) of the 4385
 simultaneous daily values of $G_{\rm Wolfer}$ and $G_{\rm Wolf}$.
One can see that Wolfer reported systematically more groups than Wolf, but the relation is significantly
 nonlinear.
In fact, as illustrated in Figure~\ref{Fig:wolf_wolfer}b, Wolfer systematically reported a few
 groups more than Wolf, at all levels of solar activity (see Section~\ref{Sec:nonlin}).
It is evident that such a nonlinear relation cannot be approximated by a single linear scaling of 1.66.
The black dash--dotted line represents the linear scaling 1.66 as proposed by SS15 and, although it reasonably
 describes low-activity periods ($G_{\rm Wolf}<4$), it progressively overestimates the
 number of groups by Wolf for high activity.
For example, for a day with 10 groups reported by Wolf, this scaling would imply 16\,--\,17 groups reported by Wolfer.
But Wolfer never reported more than 13 groups for days with $G_{\rm Wolf}=10$ (the mean for such days
 was $G_{\rm Wolfer}=11.75$).
It is therefore clear that the results obtained by applying the linear scaling ($G_{\rm Wolfer}=1.66\times G_{\rm Wolf}$)
 contradict the data.
This is also seen in Figure~\ref{Fig:wolf_wolfer}c, which shows the scatter of the Waldmeier raw daily data
 and the scaled (with a factor 1.66 as proposed by SS15) data of Wolf.
One can see that the scaling by SS15 introduces very large errors at high levels of solar activity,
 making a moderate level appearing as high.
This is a primary reason of high solar cycles claimed by SS15 and \cite{clette14} in the 18th and 19th centuries.

Figure~\ref{Fig:wolf_wolfer}d shows the relation of the simultaneous-day group numbers by the two observers,
 after the correction performed here (reduction to the reference RGO data set).
One can see that the data are nearly perfectly corrected -- the corrected data lie around the diagonal
 implying that both series report the same quantity.
The best-fit to the orange balls (the last point excluded) is $G^*_{\rm Wolfer}=(0.98\pm 0.03)\times G^*_{\rm Wolf}$,
 where the asterisks denote the values calibrated to the reference data set.
We remind the reader that Wolf and Wolfer were calibrated to the reference data set independently of each other,
 and thus this comparison provides a direct test of the validity of the method.
The ratio $G^*_{\rm Wolfer}/G^*_{\rm Wolf}$ should, of course, be unity if the calibration is done correctly:
 that this is the derived value within the uncertainty shows that both have been properly estimated.
Thus, since the ratio between the daily group numbers (for the days when both observers made observations)
 of Wolf and Wolfer is consistent with unity (\textit{i.e.} one-to-one relation) after each of them was independently
 corrected, we conclude that the method works well.
We note that this example is shown only for illustration and not used in the actual calibration.

\section{Corrected Series of Sunspot Group Numbers}

In this section we construct a composite series of sunspot group numbers for the period since 1749 using the selected observers.

\subsection{Compilation of the Corrected Series}
\label{S:compil}
For each day [$t$] we considered all those observers whose reports are available for that day.
For each such observer $i$ we took the reported $G_{{\rm obs},i}$ value for that day and corrected it to the reference
 data set using the correction matrix for that observer (as described in Section~\ref{Sec:corr}) to define the pdf
 of the corresponding values [$G_{{\rm ref}}$].
If several observers are available for the day, the corresponding pdfs were multiplied and re-normalized to unity again.
In order to avoid possible voiding of the observational days with an outlying data, we set the minimum pdf values
 to $10^{-4}$.
Then, from such composite (over all of the observers for the day) pdfs we calculated the mean daily value of $G_{\rm ref}$
 and its 68\,\% error as the standard deviation from the gathered pdf divided by $\sqrt{N}$ where $N$ is
 the number of observers for the day.

From these composite daily series we constructed a monthly composite series as a standard
 weighted average (see details given in, \textit{e.g.} \opencite{usoskin_SP_daily03}) of the composite daily values.
This series is available in the electronic supplement to the article.

\begin{figure}[t]
\centering \resizebox{\columnwidth}{!}{\includegraphics{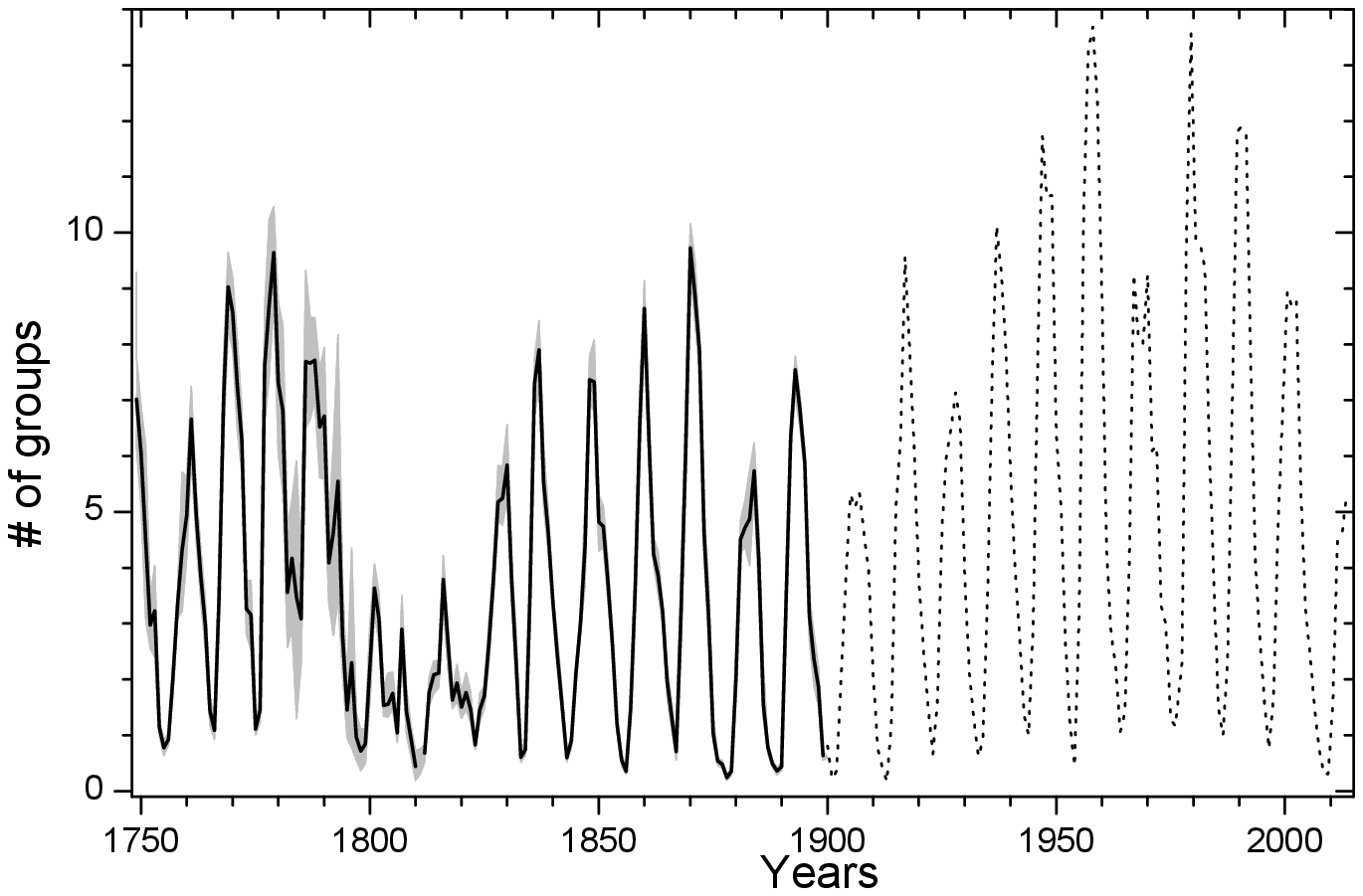}}
\caption{Annual number of sunspot groups for the period 1749\,--\,1900 (solid curve)
 normalized to the reference data set, along with 68\,\% confidence interval (grey shading).
 The reference data set of RGO group numbers for 1900-1976, extended by the SOON data 1977\,--\,2013,
 as normalized to the RGO set by Lockwood et al. (2014), is represented by the dotted curve. }
\label{Fig:all_y}
\end{figure}

The work by \cite{usoskin_SP_daily03} shows that it is not optimal to calculate the annual value of the composite series as an (arithmetic or weighted)
 average of the monthly values.
For example, if the year in question is in the rising phase of an activity cycle, the value of $G$ may increase significantly
 between January and December, rising by up to an order of magnitude.
For example, in 1867 the monthly $G$ was 0.15 in January and 1.6 in December.
If accidentally, the month of December had better coverage and a smaller error of the monthly value than January, the annual weighted
 mean would be dominated by the December value, which is obviously incorrect.
Therefore we use a Monte-Carlo procedure to calculate the annual $G$ value and its uncertainties from the daily $G$-values
 with errors:
\begin{enumerate}[i)]
\item
For each day with existing data within the given year, we took randomly
 a value of daily $G$ from the composite pdfs of corrected daily $G$-values obtained above.
\item
From these randomly taken daily $G$-values we computed the monthly values as the simple arithmetic mean.
\item
The annual value $G^{'}$ was computed as the arithmetic mean of the monthly values.
\item
Steps i)\,--\,iii) were repeated 1000 times so that a distribution of the annual $G^{'}$ values was obtained.
Finally, the mean and the 68\,\% (divided by the $\sqrt{n}$, where $n$ is the number of the months with data in the year)
 were computed from the distribution.
Errors propagate naturally and without biases in this approach.
\end{enumerate}
This annual series with 68\,\% uncertainties is shown in Figure~\ref{Fig:all_y}.
It covers the period 1749\,--\,1899, with one missing value in 1811.
To complete the series and bring it up to the present day, we use group numbers from the RGO data set
 for the period 1900\,--\,1976.
After that we used complete series by sunspot group numbers from the Solar Optical Observing Network (SOON), funded by
 the United States Air Force and NOAA with filling of some data gaps using the ``Solnechniye Danniye´´ (Solar Data, SD)
 Bulletins issued by the Main Astronomical Observatory of Russian Academy of Science (Pulkovo, St.~Petersburg, Russia).
The main SOON observations are from Boulder with support stations at Holloman, Learmonth, Palehua, Ramey, San Vito, and Mount Wilson. Unlike the RGO data, the SOON data are recorded as drawings rather than photographic plates.
We use the RGO-SOON group number intercalibration derived by \cite{lockwood_1_14}, which employs the international
 sunspot number as a spline and minimizes the difference between the means of the fit residuals for before and
 after the RGO/SOON join.

One can see that the activity remains at a moderate level in the 19th century,
 and is higher in the 18th century.
However, activity (sunspot group numbers) in both the 18th and 19th centuries remained significantly lower than the
 Modern Grand Maximum in the second half of the 20th century.

We also note that there is another source of non-linearity not accounted for here (neither was it considered
 in previous series, such as HS98, ISN or SS15).
It is related to calculation of monthly and annual values from a small number of sparse daily observations.
Under such conditions, the simple arithmetic average tends to overestimate the number of sunspots (groups)
 for active periods, if the number of daily observations per month is smaller than three (\opencite{usoskin_SP_daily03}).
The overestimate can be as much as 20\,--\,25\,\%.
This may affect the values for the 18th century where data coverage was low.
Since this effect leads to a possible overestimate of the monthly (and thus annual) values,
 it keeps the averaged series provided here as a conservative upper limit.
However, this effect does not influence the calibration and correction procedure
 which works with the original daily data.
Neither is the corrected daily series affected.
This effect will be taken into account in forthcoming studies.

\subsection{Comparison to Previous Reconstructions}

\begin{figure}[t]
\centering \resizebox{\columnwidth}{!}{\includegraphics{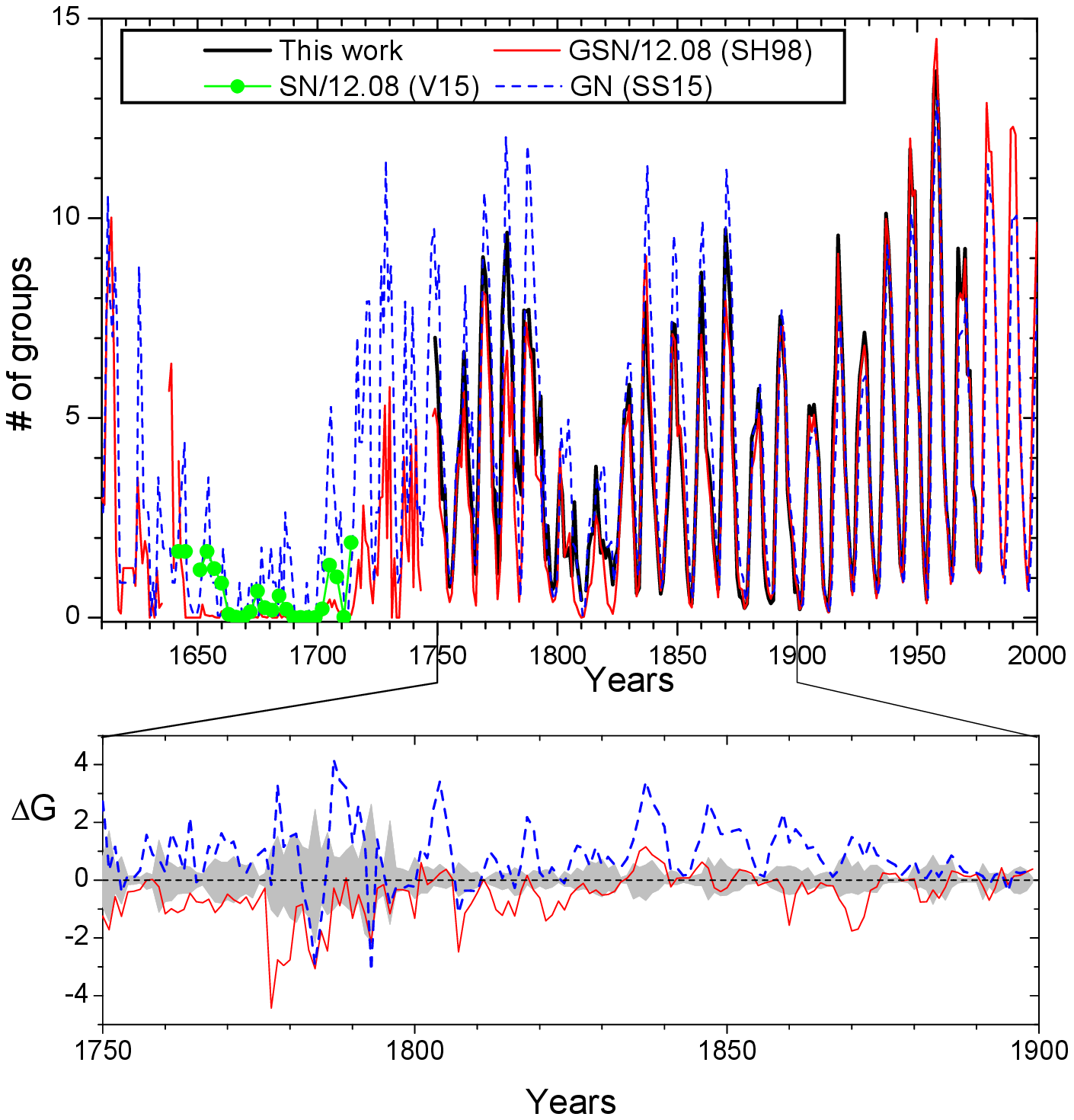}}
\caption{Comparison of different series of annual numbers of sunspot groups $G$:
 solid black curve -- the present reconstruction (identical to that shown in Figure~\ref{Fig:all_y});
 red dotted curve -- group sunspot number (HS98) divided by 12.08;
 blue dashed curve -- number of groups reconstructed by ss15;
 green dotted curve -- sunspot number divided by 12.08 around the Maunder minimum as reconstructed
   by the strict model of Vaquero et al. (2015).
  The lower panel is the difference between the two other $G$ series
   shown in the upper panel and the present result, for the period 1749-1899.
   Also plotted in the lower panel (shaded area) is the 68\,\% uncertainty estimate of the present data set.   }
\label{Fig:comp}
\end{figure}

Here we compare the new series with two earlier series of the number of sunspot groups:
 the GSN series by HS98 based on consecutive mutual calibration of observers;
 and the recent series by SS15 based on a composite of ``backbone'', ``high-low activity'',
 and ``brightest star'' methods.
In Figure~\ref{Fig:comp} we show these series along with the recent reconstruction of
 sunspot activity around the Maunder minimum using active-day statistics (\opencite{vaquero15}).

One can see that the new series is close to the GSN series by HS98, being consistent with it
 within uncertainties after $\approx$1830 and yielding cycles \#10 and 11 slightly higher than
 in the HS98 series.
The newly reconstructed cycles are significantly higher than the HS98 values before 1830.
On the other hand, the new series is  consistently lower than the values proposed by SS15, except for
 the years of the sunspot cycle minima.
This difference is mostly due to the linear regression used by SS15 to correct the Wolf-\textit{vs}-Wolfer records
 which leads to a bias, as discussed in Section~\ref{Sec:nonlin}.
Before 1830 the new series lies between the HS98 and SS15 series, on one hand being consistent with the conclusion
 by SS15 that the HS98 GSN is likely too low before the Dalton minimum, but on the other hand
 implying that the revision by SS15 is too high.
 We recall that because of the moderate to high activity underlying the reference data set,
  our reconstruction tends to overestimate reconstructed activity at earlier times with lower activity.
Note that the RGO data and the SS15 series also diverge around 1945 and tests by \cite{lockwood15a} against
 independent data indicate that this is another error in the backbone reconstruction.
The two  errors (around 1945 and the Wolf-\textit{vs}-Wolfer relationship) both act in the same direction,
 to make early group sunspot numbers too large in relation to those in the Modern Grand Maximum.

The new reconstruction suggests that the sunspot activity (quantified as the number of groups)
 was somewhat higher in the mid-18th century than it was in the mid-19th century but significantly
 lower than the Modern Grand Maximum of activity (\opencite{usoskin_PRL_03}; \opencite{solanki_nat_04}) in the
 second half of the 20th century.
The Dalton (ca. 1800) and Gleissberg (ca. 1900) lows of solar activity are clearly seen
 as the reduced magnitude of solar cycles, but they are not considered as grand minima of activity
 in contrast to the Maunder minimum (\opencite{usoskin_LR_13}).

\section{A Note on Non-linearity and the Lack of Proportionality}
\label{Sec:nonlin}
The main reason for the difference between the sunspot activity levels resulting from this work and previous
 recent reconstructions (\textit{e.g.} \opencite{clette14}; \opencite{svalgaard15})
 is that the latter were based on linear regressions between annually averaged data of individual observers, which
 can seriously distort the results.
Furthermore, SS15 assumed proportionality between the data (by forcing linear regressions through the origin and using
 only scaling correction factors without offsets) which is also incorrect and leads to possible errors which accumulate
  because the ``backbone''
  calibrations are daisy-chained (\opencite{lockwood15}).
For example the linear regression suggests a ratio of 1/0.6=1.66 between the numbers of sunspot groups
 reported by Wolf and Wolfer.
This assumes that Wolf was missing 40\,\% of all groups that would have been observed by Wolfer irrespectively of the activity level,
 \textit{viz.} a no-spot record by Wolf would correspond to no spots observed by Wolfer, but 15 groups reported by
 Wolf would correspond to 25 groups by Wolfer.
As we show here, a linear regression (proportionality) based on the annually averaged data may lead to significant biases
 leading to a heavy overestimation of the number of sunspot groups recorded by Wolf during the periods of high activity.

The relation between records of different observers is non-linear (see, \textit{e.g.}, Figure~\ref{Fig:pdf}a for Wolf).
First, it does not necessarily go through the origin (0-0 point).
Thus $G_{\rm Wolf}=0$ yields a nonzero $G_{\rm ref}$ with a mean of 0.49 and the 68\,\% range from 0 to 2 groups.
This indicates that, when Wolf reported no spots, there might have been 0\,--\,2 groups on the Sun.
This feature is totally missed when assuming a linear proportionality.
The relation between Wolf and Wolfer is quite steep (the slope of a regression forced through
 the origin is about 1.7) for low activity days
 (1\,--\,2 groups) but then the slope of the relation
 drops (see Figure~\ref{Fig:pdf}c) to almost unity for days with more than 20 groups.
It is obvious that a single correction factor is not applicable for such a relation as it assumes that an
 ``imperfect'' observer (Wolf in this case) misses the same fraction of spots irrespectively of the activity level.
However, the non-linearity of the relation is due to the fact that an observer with a poorer instrument
 or eyesight, or observing in poorer conditions,
 does not see and report the smallest spots as defined by their observational threshold $S_{\rm S}$.

\begin{figure}[t]
\centering \resizebox{\columnwidth}{!}{\includegraphics{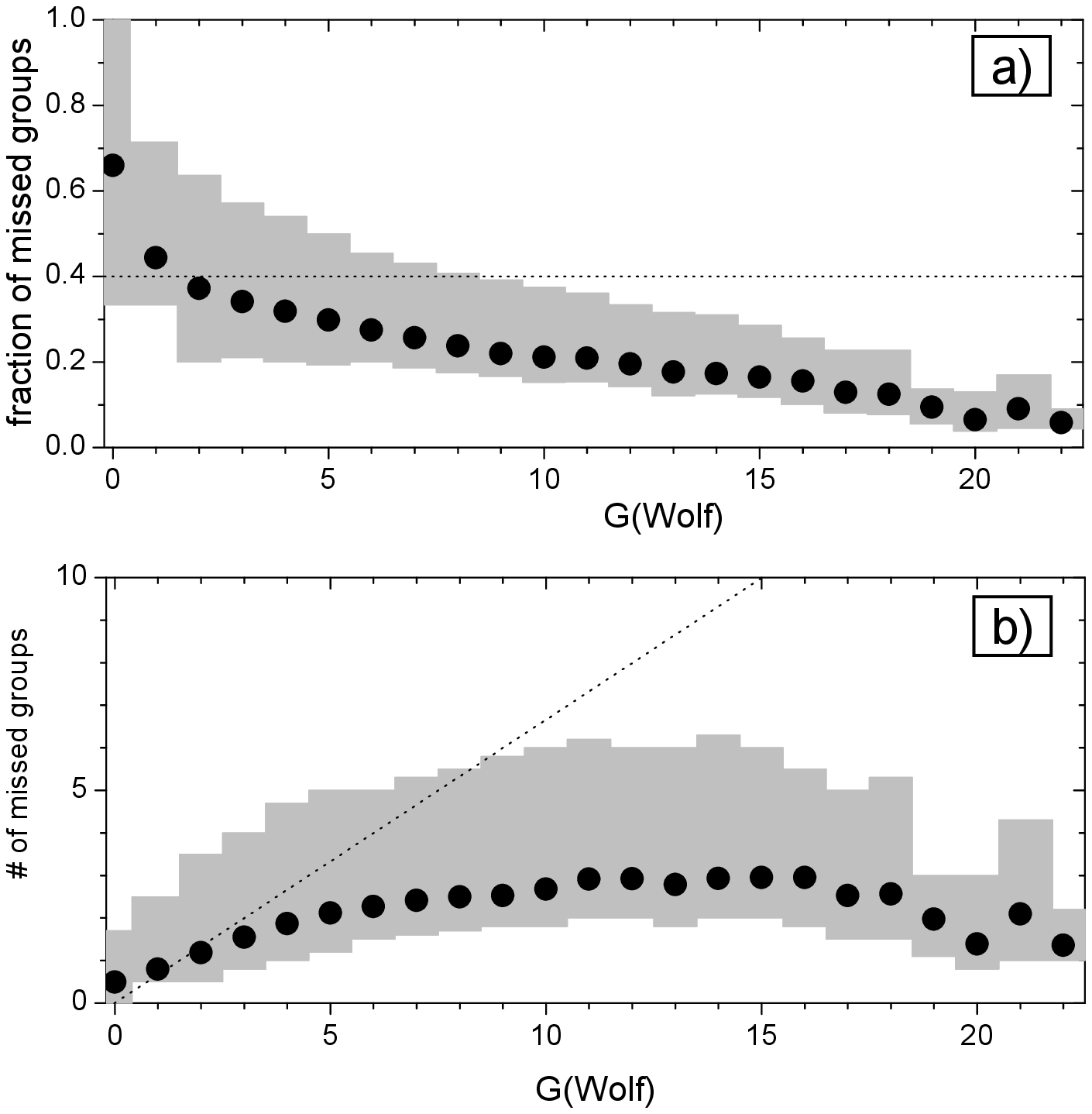}}
\caption{Assessment of small-size groups potentially missable by R. Wolf using the statistics of the
 reference data set (daily RGO sunspot groups with their size for the period 1900\,--\,1976) as shown in Figure~\ref{Fig:pdf}a.
Panel a: the fraction of the missable sunspot groups as a function of the number of sunspot groups visible for Wolf.
Panel b: the absolute number of missable sunspot groups as a function of the number of sunspot groups visible for Wolf.
 The solid dots and the grey shading depict the mean and the 68\,\% confidence intervals, respectively.
 The dotted line indicates the constant 40\,\% fraction of missed spot groups as assumed by the linear regression.  }
\label{Fig:frac}
\end{figure}

It is important that the fraction of small spots is not constant as proposed by the linear correction, but
 depends strongly on the level of solar activity.
It is usually large for a small number of groups and small for multiple groups, as illustrated in Figure~\ref{Fig:frac}
 for the case of Wolf.
While the fraction of small groups (panel a), potentially missed by Wolf, is as high as 45\,\% for the days when he
 would report only one sunspot group, it drops very quickly, so that it is only 21\,\% for days with 10 groups reported
 and 10\,\% for very active days with the number of reported groups being 16.
One can see that the assumption of a constant fraction of missed groups (the dotted line) does not describe
 the distribution, and it heavily overestimates the missed group for moderate and high activity.
From panel b one can see that, for moderate and high activity (daily $G>4$), Wolf would have been missing the same amount
 of 2\,--\,3 groups on average, irrespective of the exact number of groups he saw (\textit{cf.} Figure~\ref{Fig:wolf_wolfer}b).
The assumption of the constant fraction of missed groups (the dotted line with the slope 0.6667=0.4/0.6) is apparently
 invalid and is applicable only for low-activity days ($G<4$).
This suggests that not a multiplicative (via a scaling factor) but an additive (with an offset) correction
 would be more appropriate for moderate-high activity days.
However, the most appropriate way to correct a given observer (Wolf in this example) is to apply the correction
 matrix (Figure~\ref{Fig:pdf}) to daily group numbers observed by them as described here.

Another problem with linear regressions is that, because of the non-linearity, the averaging procedure
 is not transmissive for corrections, \textit{i.e.}, the correction of the averaged values is different from the
  average of corrected values.
Corrections must be applied to daily values using a matrix as shown in Figure\ref{Fig:pdf}, and only after that can
 the corrected values be averaged to monthly or yearly resolution, as described in Section~\ref{Sec:corr}.
Corrections applied to (annually) averaged values miss the non-linearity and, since the annual values are dominated by low and
 moderate numbers of groups, lead to an overestimate of the relation and, as a consequence, to too high solar cycles.
Moreover, the use of simple arithmetic means from sparse daily values may lead to an additional
 overestimate of the monthly (or annual) values thus distorting the relation (\opencite{usoskin_SP_daily03}, see
  discussion in Section~\ref{S:compil}).
A weighted average should be used instead.

Accordingly, the use of annual (or even monthly) averaged values for a linear scaling correction
 is not appropriate and is grossly misleading.

A different average size of sunspot groups as a function of activity level
 producing such a non-linearity is known in solar physics (\textit{e.g.}, \opencite{solanki_unruh_04}).
\textit{e.g.}, the ratio of faculae to spots changes with activity level (\textit{i.e.} the ratio of number of small to large
 magnetic flux tubes).
Of course neither faculae-to-spot area ratio nor sunspot areas directly enter into the number of sunspot
  groups or the size of groups, but they are examples of other (somewhat related)
 quantities showing a non-linear behaviour.

\section{Conclusions}

A new series of sunspot-group numbers, based on a novel method of independent calibration of the quality of
 data by different sunspot observers, is presented for the period 1749\,--\,1900.
These data have been calibrated using as the reference data set the one from the Royal Greenwich Observatory for
 the period 1900\,--\,1976,
 which is readily extended to the present day using the SOON data, forming a homogeneous series for the entire
  period since 1749\,--\,1976
 (see Figure~\ref{Fig:all_y}).
The new series is a composite of 18 individual observers (Table~\ref{Tab:Res}), each being
 independently calibrated to the reference data set using the newly developed active-day fraction method.
It is important that the new method provides independent direct calibration, using the active-day fraction
 statistics, of the observers to the reference data set in the sense of the observational threshold, thus avoiding
 consecutive error accumulation, in contrast to earlier methods.
The method includes several main steps:
\begin{enumerate}[i)]
\item
constructing calibration curves for the reference data set;
\item
 assessing the quality of individual observers in terms of the observational threshold of the
 sunspot group area [$S_{\rm S}$];
\item
 normalizing raw data by the individual observers to the reference data set;
\item
 compiling the composite series.
\end{enumerate}
The method does not use any linear regressions or other prescribed functional relations, but is, instead, based on
 a direct correction matrix and the Monte-Carlo method.
The basic assumptions of the method are:
\begin{itemize}
\item
The reference data set is of the ``perfect'' quality, and the ADF statistic for this set is representative for
 the entire period under investigation. The selection of the reference data set ensures that violations of this
 assumption may only lead to a possible overestimate of the activity.
\item
The quality of an observer is quantified as the observational threshold [$S_{\rm S}$], so that the observer misses all groups
 with an area smaller than $S_{\rm S}$ and reports all groups with an area greater than $S_{\rm S}$, and
 the quality remains constant throughout their entire period of observations, but this can be revisited
 in the future by a piece-wise calibration.
\item
The correction of an ``imperfect'' observer is based on an estimate of the number of groups (s)he would see if (s)he
 was a ``perfect'' observer
 (\textit{i.e.}, with the data quality corresponding to the reference data set).
\end{itemize}

The series presented here is a basic skeleton, or core, of the reconstruction of the number of sunspot groups,
 to which other observers with shorter
 sunspot records can and will be added later by means of direct normalization to this core series.
The raw series includes daily numbers of sunspot groups, the mean values and their 68\,\% confidence intervals,
 reduced to the reference data set, for each individual observer listed in Table~\ref{Tab:Res}.
The composite series exists, in addition to raw daily resolution, as monthly and annual averages provided
 in the Electronic Supplementary Material.

The new series is consistent with the Group Sunspot Number (\opencite{hoyt98}) after about 1830 but is
 systematically higher than that in the 18th century, implying that the sunspot activity was higher than
 proposed by HS98 before the Dalton minimum.
Conversely, the new series is significantly and systematically lower than the ``backbone'' sunspot group number SS15
 in the 19th and 18th century, implying that the ``backbone'' reconstruction grossly overestimates solar activity
 during that period.
We demonstrated that the overestimate of the ``backbone'' method was caused by the incorrect use of linear proportionality
 to normalize the observers to each other, which led to propagation and accumulation of errors.
Furthermore, the use of annual means of sparse data led to additional errors, enhancing those
 for observers active at earlier times.

The new series depicts lows of solar activity around 1800, known as the Dalton minimum, and around 1900,
 known as the Gleissberg minimum, although they are not considered as grand minima of activity.
We note that the new series provides an upper limit for the group numbers during most of the times, in particular around the
 Dalton minimum, when both the activity level and the observation density were low.
The Modern Grand Maximum of activity in the 20th century is confirmed as a unique
 event over the last 250 years.
This confirms and enhances the established pattern of the secular variability of solar activity (\textit{e.g.}
 \opencite{hathawayLR}; \opencite{usoskin_LR_13}).

\acknowledgements{
We are thankful to Rainer Arlt for the revised data of Staudacher.
Contributions from I.~Usoskin, K.~Musrsula and G.~Kovaltsov were done in the framework of the
 ReSoLVE Centre of Excellence (Academy of Finland, project no. 272157).
Work at the University of Reading is supported by the UK Science and Technology Facilities Council
 under consolidated grant number ST/M000885/1.
G.~Kovaltsov acknowledges partial support from Programme No. 7 of the Presidium RAS.
This work was partly supported by the BK21 plus program through the National Research Foundation (NRF) funded
 by the Ministry of Education of Korea.
}

\section*{Disclosure of Potential Conflicts of Interest}
The authors declare that they have no conflicts of interests.

%
%
%

\end{article}
\end{document}